\documentclass[11pt,a4paper]{article}
\usepackage{jcappub}
\usepackage{latexsym,hyperref}
\usepackage{jcappub}

\input{epsf}
\newcommand{\beq}{\begin{equation}}
\newcommand{\eeq}{\end{equation}}
\newcommand{\bea}{\begin{eqnarray}}
\newcommand{\eea}{\end{eqnarray}}
\newcommand{\bi}{\begin{itemize}}
\newcommand{\ei}{\end{itemize}}
\newcommand{\bfi}{\begin{figure}[!t]
\epsfxsize=7cm
\epsffile}
\newcommand{\bfib}{\begin{figure}[t]
\epsfxsize=9cm
\epsffile}
\newcommand{\bfig}{\begin{figure*}[t]
\epsfxsize=24cm
\epsffile}
\newcommand{\efi}{\end{figure}}
\newcommand{\efib}{\end{figure}}
\newcommand{\efig}{\end{figure*}}
\newcommand{\no}{\nonumber}

\newcommand{\bfs}{\mbox{\boldmath$s$}}
\newcommand{\bfr}{\mbox{\boldmath$r$}}
\newcommand{\bfx}{\mbox{\boldmath$x$}}
\newcommand{\bfk}{\mbox{\boldmath$k$}}
\newcommand{\bfv}{\mbox{\boldmath$v$}}
\newcommand{\bfu}{\mbox{\boldmath$u$}}

\newcommand{\bfq}{\mbox{\boldmath$q$}}
\newcommand{\bfz}{\mbox{\boldmath$z$}}

\newcommand{\hompc}{\,h\,{\rm Mpc}^{-1}}

\newcommand{\mpch}{{\rm Mpc}/h}

\title{Study on the mapping of halo clustering from real space to redshift space}
\author[a,b]{Yi Zheng,}
\author[b]{Yong-Seon Song}
\author[b,c]{and Minji Oh}

\affiliation[a]{School of Physics, Korea Institute for Advanced Study, 85 Hoegiro, Dongdaemun-gu, Seoul 130-722, Korea}
\affiliation[b]{Korea Astronomy and Space Science Institute, 776, Daedeokdae-ro, Yuseong-gu, Daejeon 34055, Republic of Korea}
\affiliation[c]{University of Science and Technology, Daejeon 34113, Korea}

\emailAdd{yizheng@kias.re.kr}
\emailAdd{ysong@kasi.re.kr}
\emailAdd{minjioh@kasi.re.kr}

\abstract{The mapping of galaxy clustering from real space to redshift space introduces the anisotropic property to the measured galaxy density power spectrum in redshift space, known as the redshift space distortion (RSD) effect. The mapping formula is intrinsically non-linear, which is complicated by the higher order polynomials due to indefinite orders of cross correlations between density and velocity fields, and the Finger--of--God (FoG) effect due to the randomness of the galaxy peculiar velocity field. In previous works, we have verified the robustness of advanced TNS mapping formula in our hybrid RSD model in dark matter case, where the halo bias models are not taken into account for the halo mapping formula in redshift space. Using 100 realizations of halo catalogs in N-body simulations, we find that our halo RSD model with the known halo bias model and the effective FoG function accurately predicts the halo power spectrum measurements, within 1$\sim$2\% accuracy up to $k\sim 0.2\hompc$, depending on different halo masses and redshifts.} 

\begin{document}
\maketitle
\flushbottom

\section{Introduction}
\label{sec:Intro}

Since the discovery of cosmic acceleration a couple of decades ago~\cite{Acceleration1,Acceleration2}, it remains as an unresolved issue to explain its  physical cause. Many theoretical models have been proposed to resolve the problem, and those can be classified into two different kinds of models. One is dark energy model in which the unknown energy component is added, and the other is modified gravity model in which the gravitational physics based upon Einstein's relativity theory is modified at cosmological scales~\cite{Lue2006,Frieman2008,Li2011,Clifton2012,Joyce2016,Koyama2016,Nojiri2017,Arun2017,Wang2017,Brax2018,Mustapha18}. Although the cosmic distances are precisely measured with multiple experiments in these days, both different models are not distinguishable exploiting those cosmic distances alone. When the growth functions are additionally observed, the mass screening effect caused by modified gravity can be probed. Then we are able to exclude either theoretical models to explain the cosmic acceleration~\cite{Amendola05,Yamamoto05,Zhang07c,Zhang08c,Linder08,Jain08,Wang08,Percival09,Song09,White09,Song10,
Wang10,Reyes10,Cai12,Gaztanaga12,Okumura15}.

We pay attention to redshift space distortion (hereafter RSD) observation of large scale structure as a promising tool to probe both cosmic expansion and growth functions simultaneously. The anisotropic features of correlation function along the line of sight are caused by peculiar motion of galaxies~\cite{Jackson72,Sargent77,Peebles80,Kaiser87,Peacock94,Ballinger96} . The careful analysis on this feature provides the information of structure formation~\cite{Peacock01,Tegmark02,Tegmark04,Samushia12,Guzzo08,Blake11b,Blake12,Marin16,Beutler16,Hector16,Satpathy16,Zhao2018,ZhengJ2018}. The same observation is also known to probe cosmic expansion history precisely through the method of baryon acoustic oscillation \cite{Seo03,Eisenstein05,Blake11a,Anderson12,Kazin14,Song14,Anderson14,Gilmarn15,Zhao16,Zhu2018,WangD2018} and Alcock-Paczynski test~\cite{APtest,LiX2016,LiX2018}. Thus we are able to observe two key cosmological observations exploiting one single observation of RSD. However the accurate and precise theoretical prediction of RSD is extremely difficult due to the contamination caused by the unknown non--linear physics, the indefinite higher order polynomials in the RSD mapping formula and the random velocity effects~\cite{Peebles80,Fisher95,Heavens98,White01,Seljak01,Kang02,Tinker06,Tinker07,
Scoccimarro04,Matsubara08a,Matsubara08b,Desjacques10,Taruya10,Taruya13,
Matsubara11,Okumura11a,Okumura11b,Sato11,Jennings11b,
Reid11,Seljak11,Okumura12,Okumura12b,Kwan12,Jennings12,Li12,
Zhangrsd,Zheng13,Ishikawa14,White15,Jennings16,Bianchi15,Bianchi16,Simpson16,Zheng16a,Song2018,Kuruvilla2018,Zhai2018,Desjacques2018}.

We have made efforts to provide more accurate RSD theoretical models recently. Being different from the conventional belief, higher order polynomials such as trispectrum related term should be included~\cite{Zheng16a} for the high precision experiment such as DESI \cite{DESI16I}, EUCLID \cite{Euclid13} and PFS \cite{PFS12}. Based upon this RSD mapping formula, the hybrid RSD model has been proposed in our previous work, which was verified using particle simulations~\cite{Song2018}. We extend our previous study into halo case in this manuscript. The halo catalogs are generated using the same particle simulations which are used to test the hybrid RSD model previously.

Halo bias is the key issue to be resolved in this manuscript. It consists of two parts, halo velocity bias $b_v$ and halo density bias $b_h$. Prior to this study, the detailed investigation on halo velocity bias has been made~\cite{Zheng14b,Junde18vb}. In \cite{Junde18vb}, authors provided an accurate fitting formula of $b_v$ at $k\leqslant 0.25\hompc$, valid for various halo mass bins at different redshifts. By calculating through this formula, we find no significant halo velocity biases for our analyzed halo catalogs at targeting scales. The $b_v$ deviates from unity at $\lesssim 1\%$ level for most cases. Thus we could approximately set $b_v=1$ and use the measured dark matter velocity field to directly represent the halo velocity field. Furthermore, in practical data analysis, galaxy distribution is fitted with the theoretical galaxy density bias model \cite{Desjacques18biasreview}. We will  apply the density bias model developed by \cite{McDonald09} to describe the halo distribution. As we will see, this model accurately predicts the two--point halo clustering from the measured dark matter clustering. Although the prediction of linear bias model fails for the higher order polynomial calculation in the RSD model, the alternative resolution of the effective FoG function is suggested to minimize this contamination to $\lesssim1\%$ at the scale of $k<0.2\hompc$. 

The paper is organized as follows. In section \ref{sec:model_sim}, we introduce our halo RSD mapping formula, halo density and velocity bias models for the test. In section \ref{sec:model_verification} we first test the halo mapping formula against halo catalogs and prove its accuracy. Then we combine the mapping formula with halo density and velocity bias models and test the halo RSD model against halo catalogs. The conclusions and discussions are given in section \ref{sec:conclusion}.

\section{Theoretical RSD model for halo clustering}
\label{sec:model_sim}
Understanding the halo clustering in redshift space is a key stepstone towards theoretically describing the observed galaxy clustering in the Universe. In our previous work~\cite{Zheng16a}, the RSD model for dark matter clustering in redshift space was studied in detail. The model has been proved to accurately reconstruct the linear growth rate within $1\%$ at $k<0.18\hompc$ for simulations of different cosmologies with different Hubble parameters \cite{Song2018}. This theoretical RSD model will be applied to the halo clustering case in this manuscript. We will describe the RSD model in this section. Besides, the halo density bias model \cite{McDonald_bias} and halo velocity bias model \cite{Junde18vb} adopted in this paper will be presented as well.

\subsection{The advanced TNS model for halos}
\label{subsec:rsd_model}
Matter distribution in the universe is inhomogeneous at small scales. The gravitational attraction arising from this inhomogeneity perturbs galaxies and causes their motions deviating from the Hubble flow. These deviations, named peculiar velocities of galaxies, disturb the galaxy redshifts and hence the galaxy distribution in redshift space in an anisotropic way. This induces anisotropic properties in galaxy clustering statistics in redshift space, such as galaxy power spectrum and bispectrum (two-- and three--point correlation function). 

If the galaxies are observed in redshift space by a sufficiently distant observer, the plane parallel approximation is applicable for transformation from real space position $\bfr$ to redshift space position $\bfs$ of a galaxy,
\beq
\label{eq:mapping}
\bfs=\bfr+\frac{\bfv \cdot \widehat{\bfz}}{a(z)H(z)}\widehat{\bfz}\,,
\eeq
where $\widehat{\bfz}$ denotes a directional unit vector of the line of sight, and $\bfv \cdot \widehat{\bfz}$ represents the physical velocity component along the $\widehat{\bfz}$ direction. The expansion scale factor and Hubble parameter at the given redshift $z$ are denoted as $a(z)$ and $H(z)$ respectively.
The mass in the given unit volume is conserved in both real and redshift spaces, which formulates the transformation between two spaces as $(1+\delta(\bfr))d^3r=(1+\delta^s(\bfs))d^3s$. The power spectrum observed in the redshift space is then given by \cite{Taruya10}, 
\begin{equation}
P^{\rm(S)}(k,\mu)=\int d^3\bfx\,e^{i\,\bfk\cdot\bfx}
\bigl\langle e^{j_1A_1}A_2A_3\bigr\rangle\,, 
\label{eq:Pkred_exact}
\end{equation}
in which we define
\begin{eqnarray}
&j_1\equiv -i\,k\mu \,,\nonumber\\
&A_1\equiv u_z(\bfr)-u_z(\bfr')\,,\nonumber\\
&A_2\equiv \delta(\bfr)+\,\nabla_zu_z(\bfr)\,,\nonumber\\
&A_3\equiv \delta(\bfr')+\,\nabla_zu_z(\bfr')\,,\nonumber
\end{eqnarray}
where $\bfx$ and $\bfu$ are defined by $\bfx\equiv\bfr-\bfr'$ and $\bfu\equiv-\bfv/(aH)$. $u_z$ is the radial direction component of $\bfu$. $\mu$ denotes the cosine of the angle between $\bfk$ and the line of sight.

Firstly we revise the dark matter RSD modelling in \cite{Zheng16a}. Eq.~(\ref{eq:Pkred_exact}) is a non-linear convolution of density and velocity field. The pairwise velocity field, $A_1$, when expanded from the exponent, produces an indefinite series of higher-order polynomials, illustrating that nonlinear mapping induces non-perturbative non-Gaussian corrections in the two--point statistics. We rewrite the integrand of eq.~(\ref{eq:Pkred_exact}) in terms of the connected moments (cumulants), using the following relation:
\bea
\langle e^{j_1A_1+j_2A_2+j_3A_3} \rangle = \exp\Bigl[\langle e^{j_1A_1+j_2A_2+j_3A_3} \rangle_c\Bigr].
\eea
Here, the ensemble $\langle\cdots\rangle_c$ stands for the cumulant. In the above, taking the derivative twice with respect to the variables $j_2$ and $j_3$ and then setting them to zero, we obtain ~\cite{Taruya10}
\bea
\langle e^{j_1A_1}A_2A_3\rangle=
\exp \left\{\langle e^{j_1A_1}\rangle_c\right\}
\left[\langle e^{j_1A_1}A_2A_3 \rangle_c+ 
\langle e^{j_1A_1}A_2\rangle_c \langle e^{j_1A_1}A_3 \rangle_c \right]. \nonumber
\eea
Then eq. (\ref{eq:Pkred_exact}) is recast as 
\bea
&P^{\rm(S)}(k,\mu)=\int d^3\bfx \,\,e^{i\bfk\cdot\bfx}\,\,
\exp \left\{\langle e^{j_1A_1}\rangle_c\right\}
\left[\langle e^{j_1A_1}A_2A_3 \rangle_c+ 
\langle e^{j_1A_1}A_2\rangle_c \langle e^{j_1A_1}A_3 \rangle_c \right].
\label{eq:Pkred_exact2}
\eea
The Finger-of-God (FoG) related term $\exp{\left\lbrace \left< e^{j_1A_1} \right> \right\rbrace}$ could be separated into two parts by its dependence on the separation vector $\bfx$ \cite{Zheng13,Zheng16a}. The ``one-point'' part $D^{\rm FoG}_{\rm 1pt}$ consists of only one-point velocity cumulants. It is moved outside the integral and represents the overall FoG term of the RSD model. The ``correlated'' part $D^{\rm FoG}_{\rm corr}$ includes auto velocity correlations. It will be Taylor expanded together with $\left[\left< e^{j_1A_1}A_2A_3\right>_c+\left< e^{j_1A_1}A_2 \right>_c\left< e^{j_1A_1}A_3 \right>_c\right]$.

RSD effect shows a 2-dimensional cylindrically symmetric anisotropy depending on $k$ and $\mu$. It is natural to perturbatively expand the formula in terms of $j_1=-ik\mu$. This perturbative approach is assumed to be applicable up to quasi--linear regime, and the truncation is made in a consistent order in terms of $j_1$. We proved in \cite{Zheng16a,Song2018} that, if we truncate the expansion at the second order of $j_1$, the dark matter mapping formula is accurate within $2\%$ at $k<0.2\hompc$, and it could reconstruct the linear growth rate within $1\%$ at $k<0.18\hompc$. The perturbed expression of $P^{\rm(S)}(k,\mu)$ has only one free parameter, the line-of-sight velocity dispersion $\sigma^2_z$, and it is given by
\bea
\label{eq:Pkred_dm}
P^{\rm (S)}(k,\mu)&=&D^{\rm FoG}(k\mu\sigma_z)P_{\rm perturbed}(k,\mu) \nonumber\\ 
&=&D^{\rm FoG}(k\mu\sigma_z)[P_{\delta\delta}+2\mu^2P_{\delta\theta}+\mu^4P_{\theta\theta} \no \\
&&\,\,\,\,\,\,\,\,\,\,\,\,\,\,\,\,\,\,\,\,\,\,\,\,\,\,\,\,\,\,\,\,\,+A(k,\mu)+B(k,\mu)+F(k,\mu)+T(k,\mu)]\,,
\eea
in which
\begin{eqnarray}
  A(k,\mu)&=& j_1\,\int d^3 x \,\,e^{i\bfk\cdot\bfx}\,\,\langle A_1A_2A_3\rangle_c\,,\nonumber\\
  B(k,\mu)&=& j_1^2\,\int d^3 x \,\,e^{i\bfk\cdot\bfx}\,\,\langle A_1A_2\rangle_c\,\langle A_1A_3\rangle_c\,,\nonumber\\
  F(k,\mu)&=& -j_1^2\,\int d^3 x \,\,e^{i\bfk\cdot\bfx}\,\,\langle u_z u_z'\rangle_c\langle A_2A_3\rangle_c\,, \nonumber\\  
  T(k,\mu)&=& \frac{1}{2} j_1^2\,\int d^3 x \,\,e^{i\bfk\cdot\bfx}\,\,\langle A_1^2A_2A_3\rangle_c\,.
 \label{eq:higher_order} 
\end{eqnarray}
The FoG term $D^{\rm FoG}$ could be approximated by a Gaussian function, whose validity was proved in simulations \cite{Zheng13,Zheng16a},
\beq
D^{\rm FoG}(k\mu\sigma_z)=\exp\left(-k^2\mu^2\sigma_z^2\right)\,.
\eeq
In this paper, the RSD model for halo clustering is formulated exploiting the same perturbative approach as for dark matter clustering. The density fluctuation $\delta_h$ and velocity divergence $\theta_h$ for halos are substituted to eq.~(\ref{eq:Pkred_dm}) and $P_h^{\rm (S)}(k,\mu)$ for halos is given by,
\bea
P_h^{\rm (S)}(k,\mu)&=&D^{\rm FoG}(k\mu\sigma_{z,h})P_{{\rm perturbed},h}(k,\mu) \no \\
&=&D^{\rm FoG}(k\mu\sigma_{z,h})[P_{\delta_h\delta_h}+2\mu^2P_{\delta_h\theta_h}+\mu^4P_{\theta_h\theta_h} \no \\
 &&\,\,\,\,\,\,\,\,\,\,\,\,\,\,\,\,\,\,\,\,\,\,\,\,\,\,\,\,\,\,\,\,\,\,\,\,\,\,+A_h(k,\mu)+B_h(k,\mu)+F_h(k,\mu)+T_h(k,\mu)]\,.
\label{eq:Pkred_halo}
\eea
We will test three functional forms to formulate $D^{\rm FoG}(k\mu\sigma_{z,h})$, namely
\begin{eqnarray}
D^{\rm FoG}(k\mu\sigma_{z,h})=\left\{ 
\begin{array}{ll}
\exp\left(-k^2\mu^2\sigma_{z,h}^2/H^2\right) & {\rm Gaussian}\,, \\
\left( 1+ k^2\mu^2\sigma_{z,h}^2/H^2 \right)^{-1} & {\rm Lorentzian}\,, \\
\left( 1+ k^2\mu^2\sigma_{z,h}^2/2H^2 \right)^{-2} & {\rm Squared~Lorentzian}\,,
\end{array}\right. 
\label{eq:Dfog}
\end{eqnarray}
where $\sigma_{z,h}$ is set to be a free parameter. This effectively allows more degrees of freedom in the FoG modelling.

The accuracy of eq.~(\ref{eq:Pkred_halo}) will be tested in section~\ref{subsec:halo_mapping}. After this, we will combine the mapping formula with robust halo bias models introduced in section~\ref{subsec:halo_density_velocity_bias}. This complete RSD model will be introduced and tested in section~\ref{subsec:halo_model_with_bias}.

\subsection{Halo density and velocity bias model}
\label{subsec:halo_density_velocity_bias}

In order to describe the relation between halo and dark matter density fields, we utilize a non-linear and non-local halo density bias model $\delta_h=\delta_h(\delta)$. This model was developed in \cite{McDonald_bias} and has been used in observational data analysis (e.g. \cite{Hector_bias,Beutler_bias}).

Following definitions of \cite{McDonald_bias} and \cite{Hector_bias}, we expand the halo density field $\delta_h(\bfx)$ in terms of the dark matter density field $\delta(\bfx)$ and its tidal tensor field $s(\bfx)$,
\bea
\label{eq:bias_x}
\delta_h(\bfx)&=&b_1\delta(\bfx)+\frac{1}{2}b_2[\delta(\bfx)^2-\sigma_2]+\frac{1}{2}b_{s2}[s(\bfx)^2-\langle s^2\rangle]\no\\
&&+{\rm higher\,\,order\,\, terms}\,.
\eea
Here $s(\bfx)=s_{ij}(\bfx)s_{ij}(\bfx)$, with $s_{ij}(\bfx)=\partial_i\partial_j\Phi(\bfx)-\delta_{ij}^{\rm Kr}\delta(\bfx)$. $\Phi(\bfx)$ is the gravitational potential. $b_1$ is the linear bias parameter, $b_2$ is the second-order local bias parameter and $b_{s2}$ is the  second-order non-local bias parameter. The terms $\sigma_2$ and $\langle s^2\rangle$ are introduced to  ensure the condition $\langle \delta_h \rangle = 0$. 

In Fourier space, eq.~(\ref{eq:bias_x}) turns out to be \cite{McDonald_bias,Hector_bias},
\bea
\label{eq:bias_k}
\delta_h(\bfk)&=&b_1\delta(\bfk)+\frac{1}{2}b_2\int \frac{d\bfq}{(2\pi)^3}\delta(\bfq)\delta(\bfk-\bfq)+\frac{1}{2}b_{s2}\int \frac{d\bfq}{(2\pi)^3}\delta(\bfq)\delta(\bfk-\bfq)S_2(\bfq,\bfk-\bfq)\no\\
&&+{\rm higher\,\, order\,\, terms}\,,
\eea
with
\beq
S_2(\bfk_1,\bfk_2) \equiv \frac{(\bfk_1\cdot\bfk_2)^2}{(k_1k_2)^2}-\frac{1}{3}.
\eeq

To be complete up to one-loop order, $P_{\delta_h\theta_h}$, the cross-power spectrum between halo density and velocity fields , and $P_{\delta_h\delta}$, the cross-power spectrum between halo density and dark matter density fields , are separately expressed as
\bea
\label{eq:pdt_bias}
P_{\delta_h\theta_h}(k)&=&b_vP_{\delta_h\theta} \no \\
&=&b_v\left(b_1P_{\delta\theta}(k)+b_2P_{b2,\theta}(k)+b_{s2}P_{bs2,\theta}(k)+b_{3\rm{nl}}\sigma_3^2(k)P^{\rm{L}}_{\rm m}(k)\right)\,, \\
\label{eq:phm_bias}
P_{\delta_h\delta}(k)&=&b_1P_{\delta\delta}(k)+b_2P_{b2,\delta}(k)+b_{s2}P_{bs2,\delta}(k)+b_{3\rm{nl}}\sigma_3^2(k)P^{\rm{L}}_{\rm m}(k)\,.
\eea
Here $\theta_h=b_v\theta$ with $b_v$ being the halo velocity bias, $b_{3\rm{nl}}$ is the third-order non-local bias parameter which contributes to the second-order corrections in the power spectrum, and $P^{\rm{L}}_{\rm m}$ is the linear dark matter power spectrum. For simplicity, we will assume that the density bias is local in Lagrangian space. This implies that the non-local bias parameters could be related to the linear bias parameter as \cite{Baldauf2012,Chan2012,Saito2014},
\bea
b_{s2}=-\frac{4}{7}(b_1-1) \,,
&\quad&
b_{3\rm{nl}}=\frac{32}{315}(b_1-1)\,.\no
\eea

In the above formulas, $P_{\delta\theta}$ and $P_{\delta\delta}$ are the dark matter density-velocity cross-power spectrum and density-density auto-power spectrum respectively. They will be directly measured from simulations in the following tests. The other power spectrum ingredients in the above formulas will be evaluated using linear perturbation theory, and their expressions are presented in Appendix \ref{appsec:highbias}.

As will be described in section~\ref{subsubsec:shotnoise}, $P_{\tilde{\delta}_h\tilde{\delta}_h}$, the measured halo density auto-power spectrum from simulations, could be decomposed into two parts,
\bea
P_{\tilde{\delta}_h\tilde{\delta}_h}&=&P_{\delta_h\delta_h}+P_{\epsilon\epsilon}\,, \no\\
P_{\delta_h\delta_h}&=&P^2_{\delta_h\delta}/P_{\delta\delta} \,.
\label{eq:pdd_bias}
\eea
Here $P_{\epsilon\epsilon}$ is the shot noise term, and it will be directly measured from simulations in section~\ref{subsubsec:shotnoise}, whose proper modelling is beyond the scope of this paper. $P_{\delta_h\delta_h}$, the determinant part of $P_{\tilde{\delta}_h\tilde{\delta}_h}$, will be modelled following eqs.~(\ref{eq:phm_bias}) and ~(\ref{eq:pdd_bias}).

For simplicity, we only consider the linear bias $b_1$ in calculating the higher order corrections ($A_h$, $B_h$, $F_h$, and $T_h$) in the model,
\bea
\label{eq:higherorder_b1_A}
A_h(k,\mu)&=&b_1^3A(k,\mu, f/b_1)\,,\\
B_h(k,\mu)&=&b_1^4B(k,\mu,f/b_1)\,, \\
F_h(k,\mu)&=&b_1^4F(k,\mu,f/b_1)\,, \\
T_h(k,\mu)&=&b_1^4T(k,\mu,f/b_1)\,.
\label{eq:higherorder_b1_T}
\eea
The detailed expressions are listed in appendix~\ref{appsec:ABFT}. The accuracy of this linear approximation and its impacts on RSD model accuracy will be studied in section~\ref{subsubsec:rsdtest_step3}

Besides $b_h$, we also consider the halo velocity bias model in this work. $b_v(k,z)$, the halo velocity bias, describes how much the halo velocity field traces that of the underlying dark matter field. $b_v$ of realistic halos, after correcting the otherwise significant sampling artifact \cite{Zhang14,Zheng14a}, was first measured in \cite{Zheng14b}. Recently, \cite{Junde18vb} developed a novel strategy to overcome the sampling artifact problem and determined this important bias parameter to $0.1-1\%$ accuracy at $k\leqslant 0.4\hompc$ and $0<z<2$, for various halo mass bins. An accurate fitting formula of $b_v(k,z)$ at $k\leqslant0.25\hompc$ was provided by \cite{Junde18vb},
\beq
b_v(k\mid M,z)\simeq 1-\left[ c_0+c_1\left( b_h(M,z)-1 \right) \right] \tilde{k}^2 \,,
\label{eq:b_v}
\eeq
where $\tilde{k}\equiv k/(\hompc)$. $c_0=-0.138\pm0.01$ and $c_1=0.186\pm0.007$ are the best fitted values found in \cite{Junde18vb}. In section~\ref{subsubsec:bv_effect} we will use eq.~(\ref{eq:b_v}) to calculate the velocity bias of halo catalogs analyzed in this paper.

\section{Verification of theoretical model}
\label{sec:model_verification}

\begin{table}[!t]
\centering
\begin{tabular}{@{}lll}
\hline\hline
parameter & physical meaning & value \\
\hline
$\Omega_m$  & present fractional matter density & $0.3132$ \\
$\Omega_{\Lambda}$ & $1-\Omega_m$ & $0.6868$ \\
$\Omega_b$ & present fractional baryon density & $0.049$\\
$h$ & $H_0/(100$~km~s$^{-1}$Mpc$^{-1})$ & $0.6731$ \\
$n_s$ & primordial power spectral index & $0.9655$ \\
$\sigma_{8}$ & r.m.s. linear density fluctuation & $0.829$ \\
\hline
$L_{\rm box}$ & simulation box size & 1890~$h^{-1}$Mpc\\
$N_{\rm p}$ & simulation particle number & $1024^3$\\
$m_{\rm p}$ & simulation particle mass & $5.46\times 10^{11}h^{-1}M_{\odot}$\\
\hline
$N_{\rm snap}$ & number of output snapshots & $13$ \\
$z_{\rm ini}$ & redshift when simulation starts & $49.0$ \\
$z_{\rm final}$ & redshift when simulation finishes & $0.0$ \\
\hline
\end{tabular}
\caption{The parameters and technical specifications of the N-body simulations for this work.}
\label{tab:simulation}
\end{table}

The same set of simulations in our previous paper~\cite{Zheng16a} is used to verify the halo RSD model here. The 100 N-body simulations were made using GADGET2~\cite{Springel05} with $L_{\rm box}=1.89\,h^{-1}$Gpc box length and $N_{\rm p}=1024^3$ particles. The volume of the simulation is close to the DESI survey volume between $z=0.8$ and $z=1.0$ \cite{DESI16I}. The cosmological parameter set for the simulations is the best fit LCDM model from PLANCK15 \cite{PLANK2015}, except the neutrino mass $m_\nu=0$. The Gaussian initial conditions of simulations are made by 2LPT code~\cite{2LPT} at $z=49$. Four snapshots at different redshifts $z=(0.0, \, 0.5, \, 0.9, \, 1.5)$ are exploited to be analyzed. The detailed simulation parameters are listed in table \ref{tab:simulation}.

The hybrid RSD model for dark matter was verified with the simulations introduced above in our previous work \cite{Zheng16a,Song2018}. In this manuscript, we test the hybrid RSD model for halos on the halo catalogs generated from the same simulations. The halo catalogs are generated by running the phase space Friends--of--Friends (FoF) ROCKSTAR halo finder~\cite{Rockstar}, with the linking length $b=0.28$. The gravitationally bounded halos are selected with three virial mass ranges of $10^{12.5}M_\odot/h-10^{13}M_\odot/h$, $10^{13}M_\odot/h-10^{13.5}M_\odot/h$, and $10^{13.5}M_\odot/h-10^{14}M_\odot/h$. The position and velocity of halos are determined by the mean position and velocity of particles at inner part of halos. The detailed specifications of the simulated halos are presented in Table~\ref{tab:halo}.

We present the stepwise test below in the order of the followings: 1) the prior test on halo RSD mapping formulation, 2) test on the bias model, and 3) verification of halo RSD formulation combined with bias modelling.

\begin{table}[!t]
\begin{center}
\begin{tabular}{r|c|c|c|cccccc}\hline\hline
Set ID & $\log M$ range & $N_h/10^5$ &$n_h$ &$b_1$\\\hline
\textit{LB}($z=0.0$) & $12.5$-$13.0$ & 37.6 & 5.6 & 1.05\\
                 $z=0.5$\ \ &  $12.5$-$13.0$ & 35.3 & 5.2 & 1.38 \\
                 $z=0.9$\ \ & $12.5$-$13.0$ & 29.9& 4.4 & 1.80 \\
                 $z=1.5$\ \ & $12.5$-$13.0$ & 19.4 & 2.9 & 2.61\\ \hline\hline
\textit{MB}($z=0.0$) & $13.0$-$13.5$ & 19.8 & 2.9 & 1.29 \\
                 $z=0.5$\ \  &  $13.0$-$13.5$ & 16.3 & 2.4 & 1.80 \\
                 $z=0.9$\ \  &  $13.0$-$13.5$ & 12.1& 1.8 & 2.37 \\
                 $z=1.5$\ \  &  $13.0$-$13.5$ & 6.1 & 1.0 & 3.54\\\hline\hline
\textit{HB}($z=0.0$) & $13.5$-$14.0$ & 6.7& 1.0 & 1.74\\
                 $z=0.5$\ \  &  $13.5$-$14.0$ & 4.6 & 0.7 & 2.52 \\
                 $z=0.9$\ \  &  $13.5$-$14.0 $ & 2.7 & 0.4 & 3.42 \\
                 $z=1.5$\ \  &  $13.5$-$14.0$ & 0.8 & 0.1 & 5.19\\\hline\hline
\end{tabular}
\end{center}
\caption{Three sets of halo mass bins. \textit{LB}: low mass bin, \textit{MB}: middle mass bin, \textit{HB}: high mass bin. The logarithmic mass unit is $M_\odot/h$ and the halo number density $n_h$ has unit of $10^{-4}(\mpch)^{-3}$.  $N_h$ is the total halo number in a halo mass bin. The linear density bias $b_1$ is fitted in figure~\ref{fig:shotnoise}. }
\label{tab:halo}

\end{table}


\subsection{The prior test on halo RSD mapping formula}
\label{subsec:halo_mapping}

Before the halo RSD model is fully verified, an intermediate step is introduced in this subsection. It is assumed that both density fluctuations and peculiar velocities of halos in real space are known. Then we are able to test the mapping formulation of halos itself.

\begin{figure}[!t]
\centering
\includegraphics[width=0.5\textwidth]{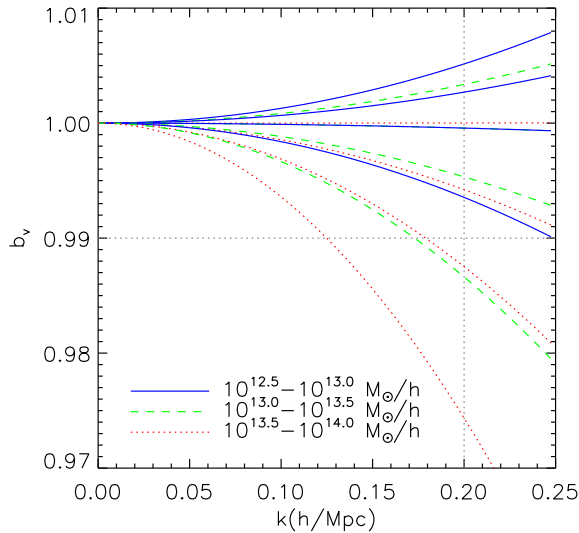}
\caption{The velocity bias of each halo mass bin calculated from eq.~(\ref{eq:b_v}). Curves with different colors and line-styles represent $b_v$ of different halo mass bins, and from top to bottom, lines correspond to $b_v$ calculations at $z=(0.0,\,0.5,\,0.9,\,1.5)$ respectively. In particular, low mass bin at $z=0.9$ and middle mass bin at $z=0.5$ have the same $b_1$, thus their $b_v$'s overlap in the figure.}
\label{fig:b_v}
\end{figure}

\begin{figure}[!t]
\centering
\includegraphics[width=0.95\textwidth]{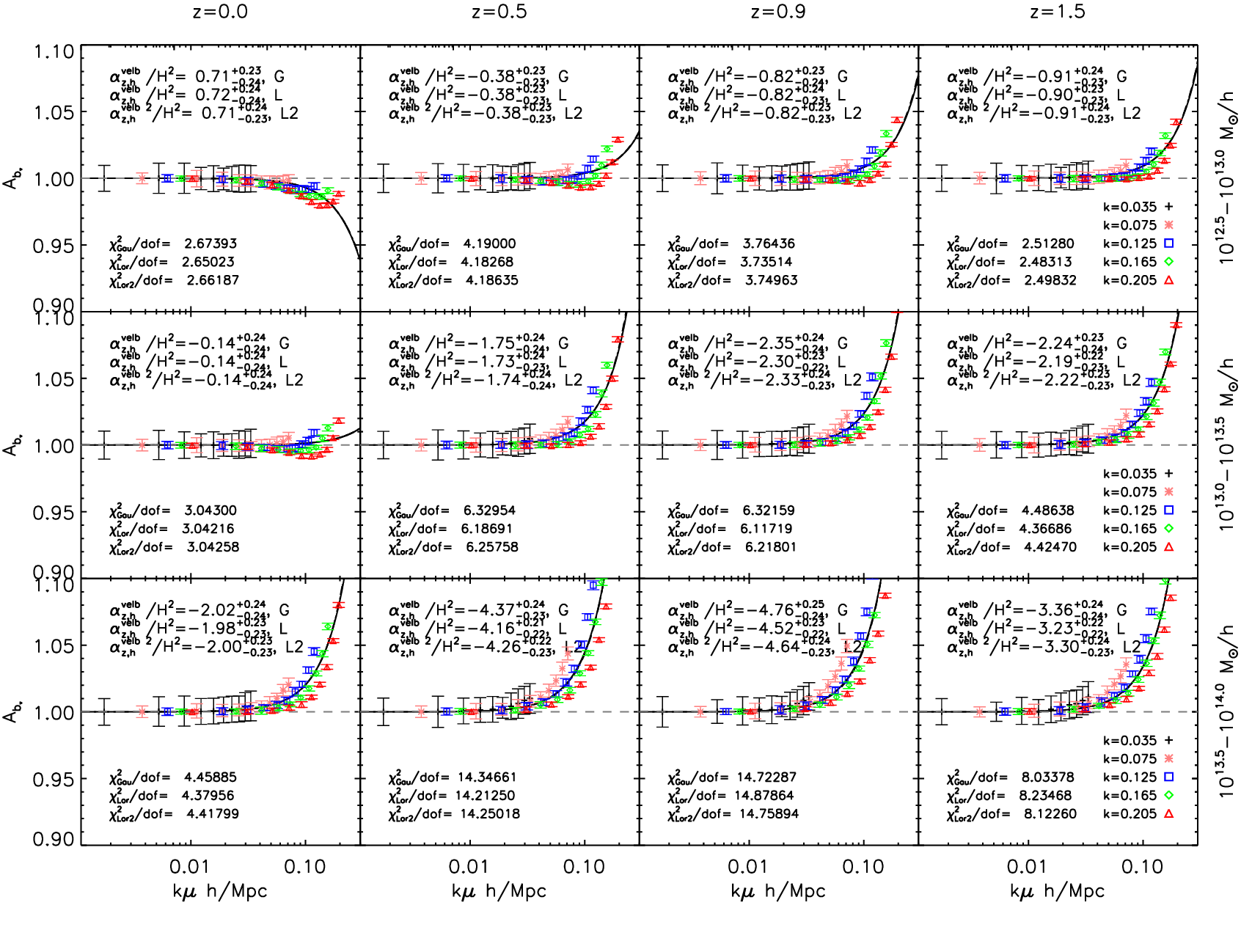}
\caption{The measurements of $A_{b_v}$ from eq. (\ref{eq:A_bv}) are shown. The data points are the averaged values from 50 simulations. The error bars come from the standard errors of the mean of measured $P_h^{\rm (S)}(k,\mu)$. Panels from top to bottom show results for halos with mass $10^{12.5}-10^{13.0}M_\odot/h$, $10^{13.0}-10^{13.5}M_\odot/h$ and $10^{13.5}-10^{14.0}M_\odot/h$ respectively. From left to right, columns represent results at $z=(0.0,\,0.5,\,0.9,\,1.5)$. The theoretical lines are model predictions adopting different FoG terms (eq.~(\ref{eq:Dfog})) fitted with data at $k\sim0.035-0.205\hompc$. Solid lines represent squared Gaussian fitting formula. Dashed lines represent Squared Lorentzian fitting formula. Dot-dashed lines represent Lorentzian fitting formula. The fitted ${\sigma_{z,h}^{\rm velb}}^2$ and $\chi^2$/dof of three FoG functions are shown in the legend of each panel.}
\label{fig:A_bv}
\end{figure}

The anisotropic halo density power spectrum $P_h^{\rm (S)}(k,\mu)$ in redshift space is measured to begin with. The nearest grid point (NGP) is used to sample the redshift space halo density field $\tilde \delta_h^s$ on $512^3$ regular grid points in configuration space. This measured $\tilde \delta_h^s(\bfx)$ field is transformed to $\tilde \delta_h^s(\bfk)$ in Fourier space by Fast Fourier Transform (FFT). The power spectrum $P_h^{\rm (S)}$ in redshift space is computed by $P_h^{\rm (S)}(k,\mu)=\left\langle\tilde \delta_h^s(\bfk)\tilde \delta_h^s(-\bfk)\right\rangle$ using the transformed $\delta_h^s(\bfk)$ in Fourier space.

\subsubsection{The effect of halo velocity bias}
\label{subsubsec:bv_effect}
The measurements of real space power spectra and higher order polynomials in eq.~(\ref{eq:Pkred_halo}) require the sampling of volume-weighted halo velocity field. However, halos are sparsely and inhomogeneously distributed, thus the sampled velocity field is severely contaminated by the sampling artifact even at the linear scales of $k\sim 0.1\hompc$ \cite{Zhang14,Zheng14a}. As shown in eq.~(\ref{eq:b_v}), \cite{Junde18vb} developed a novel strategy to overcome this sampling artifact problem and proposed an accurate fitting formula for $b_v(k,z)$ at $k\leqslant0.25\hompc$. We estimate the halo velocity bias by substituting the linear halo density biases in Table~\ref{tab:halo} to eq.~(\ref{eq:b_v}). Figure~\ref{fig:b_v} shows the calculated velocity biases. In figure~\ref{fig:b_v}, curves  with different colors and line-styles represent $b_v$ of different halo mass bins, and from top to bottom, lines correspond to $b_v$ calculations at $z=(0.0,\,0.5,\,0.9,\,1.5)$ respectively. Most of the estimated velocity biases have no bigger than $1\%$ deviation from unity at $k\lesssim 0.2\hompc$. Thus  it is safe to set $b_v=1$ for theoretical calculations of $P_{\delta_h\theta_h}$ and $P_{\theta_h\theta_h}$ in eq. (\ref{eq:Pkred_halo}) hereafter.



However, the effect of $b_v$ on higher order terms is a bit more complicated. Higher order terms are expressed as integrals of power spectra, bispectra, trispectra, etc. The integration interval includes small scales of $k>0.2\hompc$, where the halo velocity bias could significantly deviate from unity. To this point, it is not straightforward to predict the actual influence of halo velocity bias on RSD effect. To overcome this complexity, we construct an estimator, analogous to the estimator for studying the multi-streaming effect in \cite{Zheng16c}, that is
\beq
A_{b_v} \equiv \frac{P_h^{\rm(S)}(k,\mu)}{P_{h,v_{DM}}^{\rm(S)}(k,\mu)} \, ,
\label{eq:A_bv}
\eeq
in which 
\beq
\label{eq:denominator}
P_{h,v_{DM}}^{\rm(S)}(k,\mu) \equiv \int d^3\bfx\,e^{i\,\bfk\cdot\bfx} \bigl\langle e^{j_1(u_z(\bfr)-u_z(\bfr'))}(\delta_h(\bfr)+\nabla_z u_z(\bfr))(\delta_h(\bfr')+\nabla_z u_z(\bfr'))\bigr\rangle\,.
\eeq
In this estimator, $P_{h,v_{DM}}^{\rm(S)}(k,\mu)$ is evaluated by substituting the sampled halo density field and dark matter velocity field into eq. (\ref{eq:denominator}). Referring to eq. (\ref{eq:Pkred_exact}), it is easy to find that $A_{b_v}$ fully quantifies the influence of halo velocity bias on the RSD effect, except that the small scale (smaller than grid size) dark matter velocity dispersion is omitted in calculating $P_{h,v_{DM}}^{\rm(S)}(k,\mu)$. In \cite{Zheng16c}, this grid effect is used to study the multi-streaming effect of dark matter RSD, while for the halo case here, since halo velocity averages over the dark matter velocity inside a halo, the multi-streaming effect is already very much suppressed in $P_{h,v_{DM}}^{\rm(S)}(k,\mu)$. Therefore, under the assumption of negligible multi-streaming effect, we consider $A_{b_v}$ as a robust estimator in quantifying the systematic effect of halo velocity bias, and the results are shown in figure~\ref{fig:A_bv}.

In general, for the halo mass bins whose velocity biases are lower than unity, we see that $A_{b_v}$ is greater than 1, indicating that $P_h^{\rm (S)}$ is less damped than that induced by dark matter velocity field. The systematic error, as quantified by $A_{b_v}$, could be as large as $10\%$ for large halos at $k\mu = 0.2\hompc$. Figure~\ref{fig:A_bv} falsifies the usual assumption that we could represent halo velocity field by dark matter velocity field. Furthermore, $A_{b_v}$ is not a pure function of $k\mu$. It shows obvious $k$ and $\mu$ dependence separately. The detailed modelling of $A_{b_v}$ is beyond this paper's scope. We will assume $A_{b_v}$ is a only function of $k\mu$ and approximate it with three FoG functional forms. 
\begin{eqnarray}
A_{b_v}=\left\{ 
\begin{array}{ll}
\exp\left(-k^2\mu^2\alpha^{\rm velb}_{z,h}/H^2\right) & {\rm Gaussian}\,, \\
\left( 1+ k^2\mu^2\alpha^{\rm velb}_{z,h}/H^2 \right)^{-1} & {\rm Lorentzian}\,, \\
\left( 1+ k^2\mu^2\alpha^{\rm velb}_{z,h}/2H^2 \right)^{-2} & {\rm Squared~Lorentzian}\,,
\end{array}\right. 
\end{eqnarray}
The fitted $\alpha^{\rm velb}_{z,h}$ is shown on the top left corner of each panel.

\subsubsection{The mapping formula accuracy test}
\label{subsubsec:mapping_test}
\begin{figure}[!t]
\centering
\includegraphics[width=0.95\textwidth]{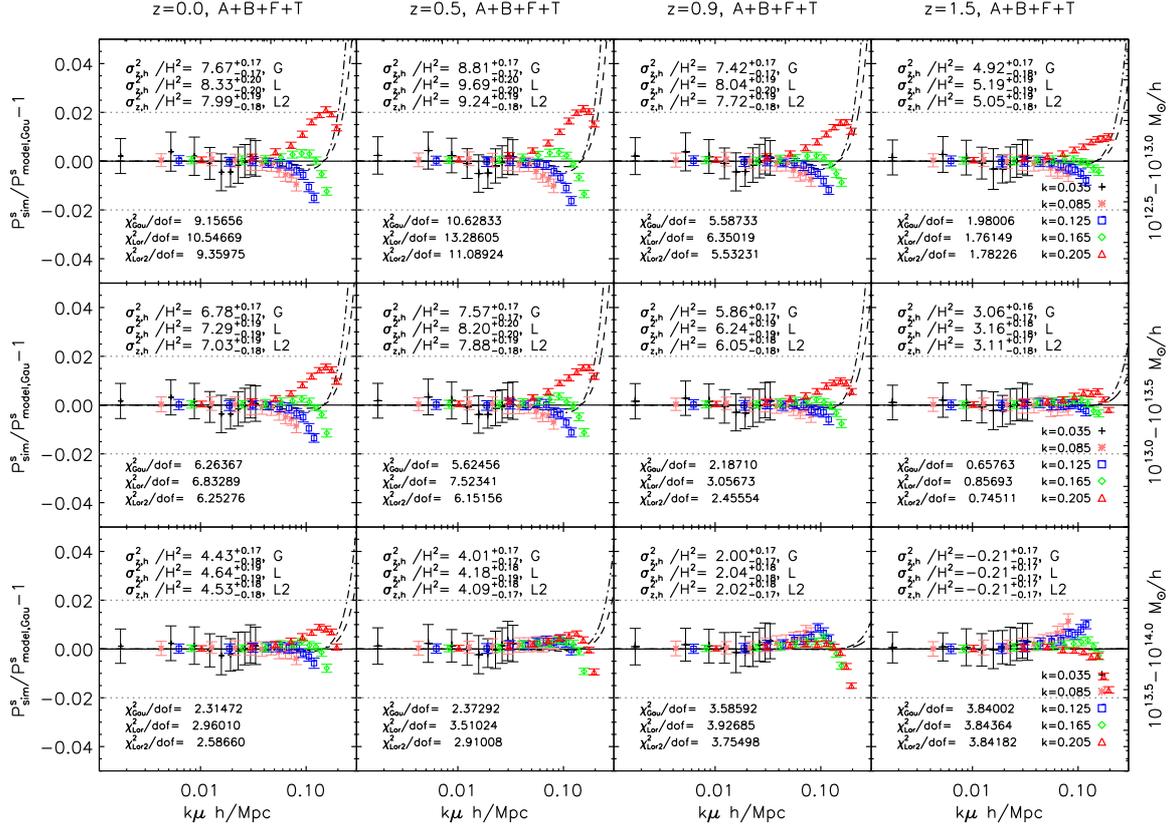}
\caption{The differences between the measured $P_h^{(S)}(k,\mu)$ and the fitted $P_h^{\rm (S)}(k,\mu)$ from eq.~(\ref{eq:Pkred_halo}). The RSD model includes $A+B+F+T$ terms. The data points are the averaged values from 100 simulations. The error bars come from the standard errors of the mean of measured $P_h^{\rm (S)}(k,\mu)$. Panels from top to bottom show results for halos with mass $10^{12.5}-10^{13.0}M_\odot/h$, $10^{13.0}-10^{13.5}M_\odot/h$ and $10^{13.5}-10^{14.0}M_\odot/h$ respectively. From left to right, columns represent results at $z=(0.0,\,0.5,\,0.9,\,1.5)$. The theoretical lines are model predictions adopting different FoG terms (eq.~(\ref{eq:Dfog})) fitted with data at $k\sim0.035-0.205\hompc$. Solid lines represent squared Gaussian fitting formula. Dashed lines represent Squared Lorentzian fitting formula. Dot-dashed lines represent Lorentzian fitting formula. The fitted $\sigma_{z,h}^2$ and $\chi^2$/dof of three FoG functions are shown in the legend of each panel.}
\label{fig:fog_haloden_frac}
\end{figure}
\begin{figure}[!th]
\centering
\includegraphics[width=0.95\textwidth]{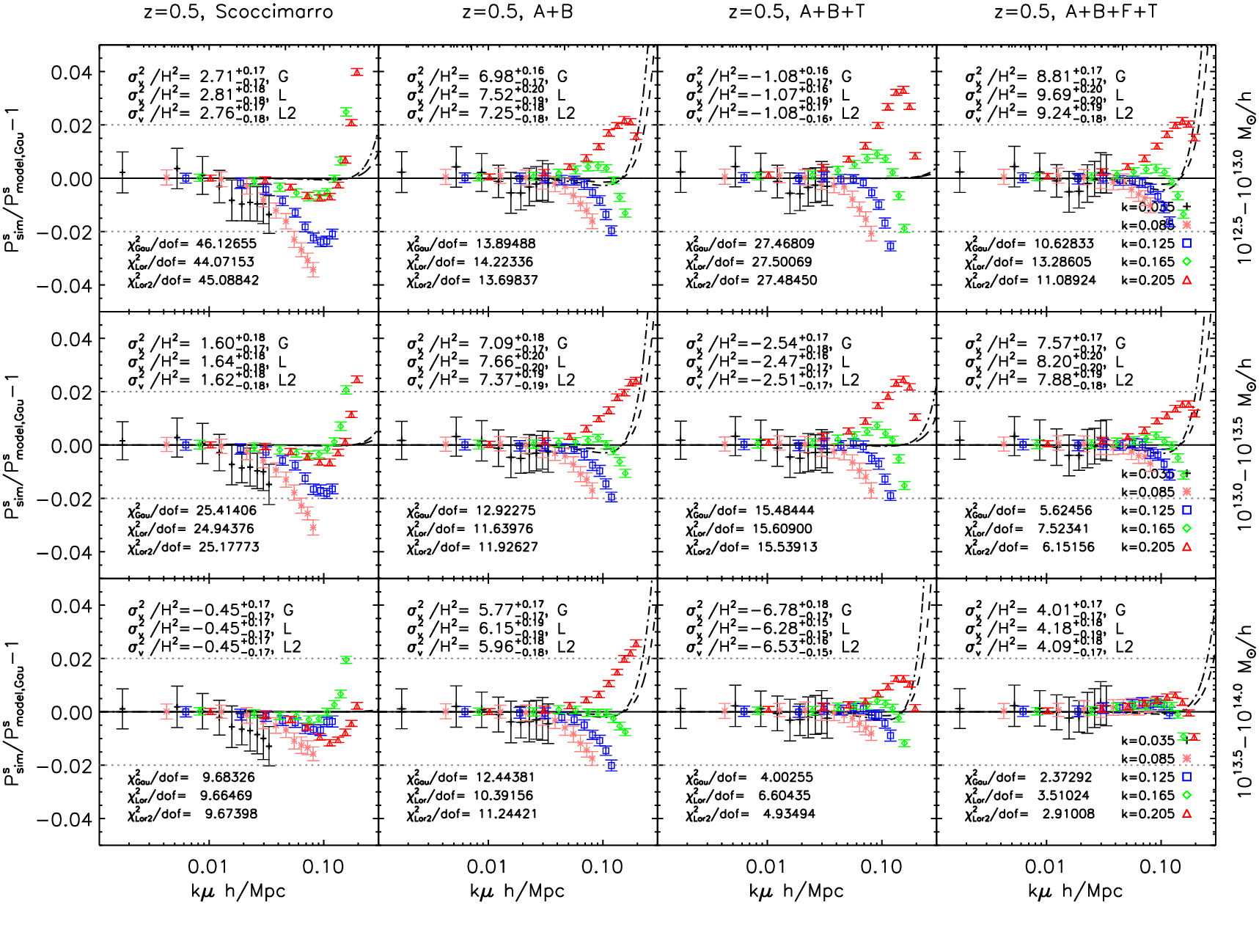}
\caption{Similar to figure~\ref{fig:fog_haloden_frac}, but we compare the RSD model including $A+B+F+T$ terms with the one without higher order terms (Scoccimarro model) and those including only $A+B$ and $A+B+T$ terms at $z=0.5$.}
\label{fig:fog_haloden_frac_ab}
\end{figure}

Since $A_{b_v}$ fully quantifies the halo velocity bias effect, from now on we replace $\theta_h$ with $\theta$, the volume--weighted dark matter velocity fields.  The dark matter velocity fields are computed by the nearest particle method (NP method)~\cite{Zheng13}. In other words, the velocity of the nearest dark matter particle to each grid is assigned to this grid for computing $\theta$. The dark matter particles have much higher number density than that of halos. The sampling artifact is controlled to be less than $1\%$ in our simulations at $k\leqslant0.2\hompc$ \cite{Zheng16c}. In practice, $P_{\delta_h\theta_h}$ and $P_{\theta_h\theta_h}$ are replaced by $P_{\delta_h\theta}$ and $P_{\theta\theta}$ in our test. The same treatment is applied to higher order polynomial measurements, which are computed by combining fields such as of $u_z(\bfr)$, $\nabla_zu_z(\bfr)$, $\delta_h u_z(\bfr)$ and $u_z\nabla_z u_z(\bfr)$. The detailed methodology of higher order polynomial calculation is explained in~\cite{Zheng16a} with dark matter particles as an example. The key strategy is that we compute these various field combinations in the configuration space, and transform them into the Fourier space, where we complete all two--point statistical measurements and derive $P_{{\rm perturbed},h}$. With the measured $P_h^{\rm (S)}(k,\mu)$ and $P_{{\rm perturbed},h}$ on hand, we could test the accuracy of eq.~(\ref{eq:Pkred_halo}) by fitting the FoG term through the least-$\chi^2$ method. The fitting range of $k$ is chosen to be $0.035-0.205\hompc$, with bin size $\Delta k=0.01\hompc$. 

As pointed out in \cite{Zhangrsd,Zheng16a}, the FoG term is an exponential function whose index contains indefinite orders of terms to be formulated in the closed form. It was verified in \cite{Zheng13,Zheng16a} that this FoG term could be effectively formulated with a simple Gaussian function, as the leading order term dominates in our interesting range of scales. It was also verified in \cite{Zheng13} that the bulk flow component of peculiar velocity field dominates the FoG term, and figure~\ref{fig:b_v} indicates that halo and dark matter have similar bulk flow components. Hence we expect the Gaussian function will still be a good FoG approximation, which is proved in figure~\ref{fig:fog_haloden_frac}.

In our test, we define $D^{\rm FoG}_{\rm res}$, the measured residual FoG term, as $P_h^{\rm (S)}(k,\mu)/P_{{\rm perturbed},h}$. If the perturbed term $P_{{\rm perturbed},h}$ is correctly estimated, the residual FoG will be well represented by the single Gaussian function in terms of $k\mu$, regardless of different $k$. Figure~\ref{fig:fog_haloden_frac} shows the fractional difference between the measured $P_h^{\rm (S)}$ and our model with Gaussian FoG function for three halo mass bins at 4 redshifts. It is effectively the fractional difference between $D^{\rm FoG}_{\rm res}$ and the best fitted Gaussian FoG function. Our RSD mapping formula is proven to be accurate within $1\sim2\%$ at $k\lesssim0.2\hompc$, depending on halo bins with different masses and redshifts. In addition, we compare the best fitted Lorentzian FoG model (dashed line) and Lorentzian2 FoG model (dot-dashed line) with the best fitted Gaussian model.  The corresponding reduced $\chi^2$/dof and best fitted velocity dispersion $\sigma^2_{z,h}$ of three FoG forms listed in each panel. With different best fitted $\sigma^2_{z,h}$ three FoG functions do not show much difference and Gaussian function is verified to be a good FoG approximation.

We further compare the effects of different higher order term combinations in our model in figure~\ref{fig:fog_haloden_frac_ab}. We present the various combinations of higher order polynomials from the left to the right. ``Scoccimarro'' denotes the model proposed in \cite{Scoccimarro04} where no higher order polynomials are considered, as $P_{{\rm perturbed},h}=P_{\delta_h\delta_h}+2\mu^2P_{\delta_h\theta_h}+\mu^4P_{\theta_h\theta_h}$. It certainly unfits to the measurements. Conventionally, only $A+B$ combinations are adapted for most RSD data analysis, which shows the good fit only at $k\lesssim0.1\hompc$. When all combinations of $A_h+B_h+F_h+T_h$ are used, the fitting has the lowest $\chi^2$/dof and the residual deviates from the simple Gaussian function within $1\sim2\%$ accuracy at $k\lesssim0.2\hompc$.


\subsubsection{The fitted velocity dispersion}
\label{subsubsec:fitted_sigmav}

\begin{figure}[!t]
\centering
\includegraphics[width=0.5\textwidth]{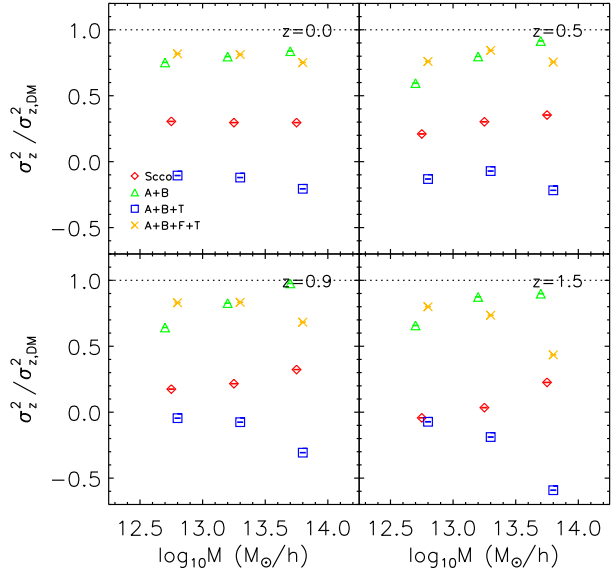}
\caption{The ratios $\sigma_z/\sigma_{z,DM}$ of different FoG functions, halo mass bins, and redshifts. $\sigma_z^2=\sigma_{z,h}^2-{\sigma_{z,h}^{\rm velb}}^2$ is the fitted velocity dispersion after correcting the halo velocity bias effect in figure~\ref{fig:A_bv}.}
\label{fig:sigmav}
\end{figure}
In eq. (\ref{eq:A_bv}), we define an estimator, $A_{v_b}$, which fully quantifies the systematic from the halo velocity bias.
Therefore, the redshift space 2D halo density power spectrum, $P_h^{\rm (S)}$, could be expressed as
\bea
P_h^{\rm(S)}(k,\mu) &=& A_{b_v} P_{h,v_{\rm DM}}^{\rm(S)}(k,\mu) \\ \no
&=&\exp(-k^2\mu^2\alpha^{\rm velb}_{z,h}/H^2) \\ \no
&&\times\int d^3\bfx\,e^{i\,\bfk\cdot\bfx} \bigl\langle e^{j_1(u_z(\bfr)-u_z(\bfr'))}(\delta_h(\bfr)+\nabla_z u_z(\bfr))(\delta_h(\bfr')+\nabla_z u_z(\bfr'))\bigr\rangle\,. \no
\eea
Here we describe $A_{b_v}$ by a Gaussian function as an example, which is supported by figure \ref{fig:A_bv}. In figure~\ref{fig:A_bv} it also shows that $A_{b_v}$ could be either higher or lower than unity, meaning that $\alpha^{\rm velb}_{z,h}$ could be either negative or positive. The full understanding of $A_{b_v}$ calls for understanding how halo velocity bias plays its role in RSD effect, and we would like to address this issue in the future.

We go on the analyze $P_{h,v_{\rm DM}}^{\rm(S)}(k,\mu)$. It is now fully constructed by dark matter velocity field, and its FoG effect is naturally controlled by the dark matter velocity dispersion.
\bea
P_h^{\rm(S)}(k,\mu) &=& A_{b_v} P_{h,v_{\rm DM}}^{\rm(S)}(k,\mu) \\ \no
&=&\exp(-k^2\mu^2\alpha^{\rm velb}_{z,h}/H^2)\exp(-k^2\mu^2\sigma^2_{\rm DM}/H^2)P_{{\rm perturbed},h} \\ \no 
&=&\exp(-k^2\mu^2\sigma_{z,h}^2/H^2)P_{{\rm perturbed},h} \,. \no
\eea
As a result, the fitted velocity dispersion, $\sigma_{z,h}^2$, has two components,  $\sigma_{z,h}^2 = \alpha^{\rm velb}_{z,h} +\sigma^2_{\rm DM} $. It is $\sigma^2_{z} = \sigma_{z,h}^2 - \alpha^{\rm DM}_{z,h}$ which will be fairly compared to the measured dark matter velocity dispersion in figure~\ref{fig:sigmav}.

The ratios of $\sigma^2_{z}$ and $\sigma^2_{z,\rm DM}$ are shown in figure~\ref{fig:sigmav}. Different colors and symbols represent different combinations of higher order terms. All ratios are all below unity, in particular, $A+B+T$ model results in a negative $\sigma^2_{z}$ and invalidates it as a robust RSD model. $\sigma^2_z/\sigma^2_{z,\rm DM} \approx 1$ is expected if the expansion in terms of $j_1$ converges and the convergence will possibly be improved if we include higher order of $j_1$ terms, e.g. $j_1^3$ and $j_1^4$ terms. The results will be published in a companion paper. 

\subsection{Test of the halo RSD model with full bias models}
\label{subsec:halo_model_with_bias}

We have verified that the measured anisotropic halo power spectrum is well fitted with the halo RSD mapping formulation itself. In practice, halo or galaxy density fields are not directly predictable from the given cosmological models. In this subsection, we continue our test in the previous subsection with halo bias models in addition to the measured dark matter fields.

\subsubsection{Shot noise term in halo density auto-power spectrum}
\label{subsubsec:shotnoise}

\begin{figure}[!t]
\centering
\includegraphics[width=0.49\columnwidth]{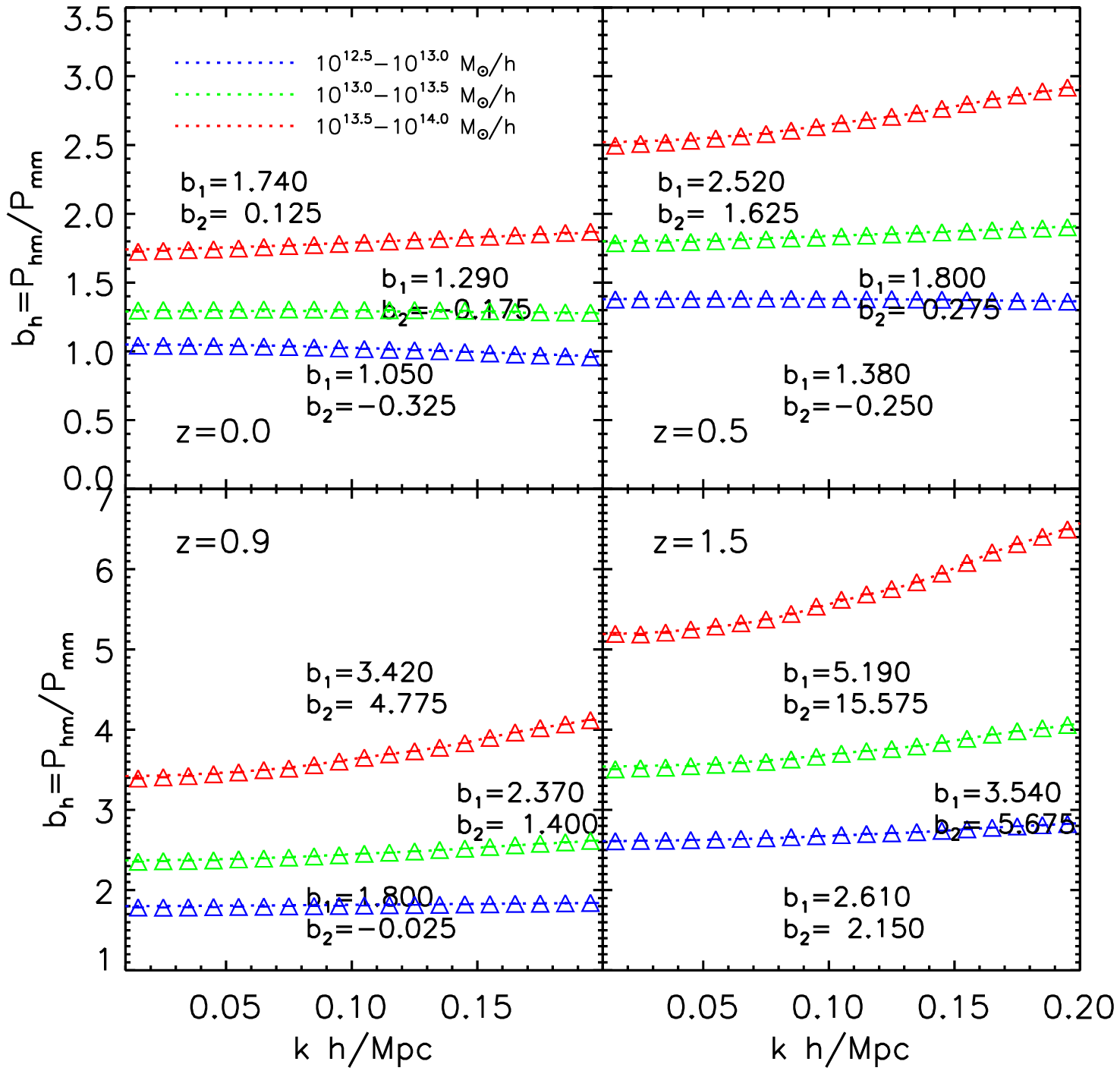}
\hfill
\includegraphics[width=0.49\columnwidth]{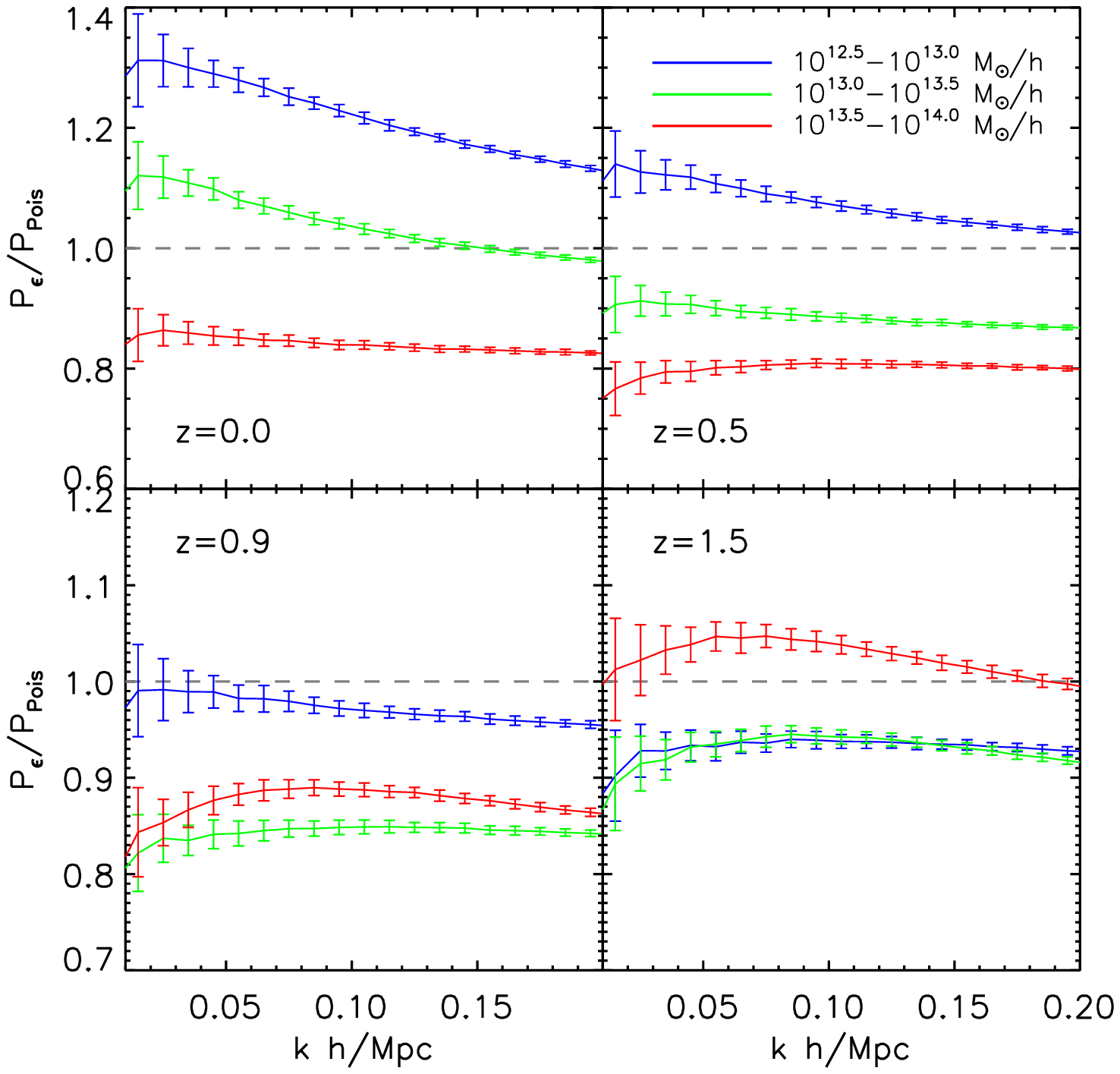}
\caption{{\it Left:} The triangles with error bars are measured halo density bias $b_h=P_{\delta_h\delta}/P_{\delta\delta}$ from simulations. The dotted lines are fitted bias models from eq.~(\ref{eq:pdd_bias}) with corresponding fitted $b_1$ and $b_2$ also shown beside lines. {\it Right:} The calculated stochastic terms divided by the corresponding Poisson noise term. The error bars are standard errors of the measurements. The scale dependent sub- and super-Poissonian property of the halo density stochastic terms are clearly visible. }
\label{fig:shotnoise}
\end{figure}

We have halos with limited number density, which undergo through nonlinear evolution and have discrete and stochastic distribution. The resultant shot noise or stochastic influence needs to be controlled. For a more precise test, this uncertainty due to the stochastic term needs to be removed from the measured halo density fields. 

If halos are distributed by Poisson process, the shot noise term is a constant, simply given by the inverse of halo number density, $1/\bar{n}_h$.  However in reality, the measured shot noise at large scales exhibits the scale dependence due to the halo exclusion and  nonlinear enhancement of clustering outside the exclusion scale~\cite{Baldauf13}. This scale dependent shot noise term should be taken into account in our test. 

First we describe the way to calculate the stochastic term. The measured halo density fluctuations $\tilde{\delta}_h(\bfk)$ can be decomposed into two components, the determinent halo density fluctuations $\delta_h(\bfk)$ and the stochastic uncertainty $\epsilon(\bfk)$~\cite{Seljak09,Nishimichi11},
\bea
\tilde{\delta}_h(\bfk)=\delta_h(\bfk)+\epsilon(\bfk)=b(k)\delta(\bfk)+\epsilon(\bfk)\,.
\eea
By definition, the dark matter density fluctuations $\delta(\bfk)$ does not correlate with the halo stochastic noise field $\epsilon(\bfk)$, $<\epsilon(\bfk)\delta^\ast(\bfk)>=0$. Here $b(k)$ is the deterministic halo density bias, and it can be measured by
\bea
\label{eq:bias}
b(k)&=&P_{\tilde{\delta}_h\delta}(k)/P_{\delta\delta}(k)\,. \no 
\eea
The measured halo density bias $b(k)$ are plotted as data points in the left panel of Fig.~\ref{fig:shotnoise}. The error bars are estimated by 100 realizations of simulations. Generally it shows that, the bias becomes more non-linear, and its scale dependence becomes larger with increasing redshift and halo mass.

Next, the stochastic power spectrum $P_{\epsilon\epsilon}$ can be calculated by
\bea
P_{\epsilon\epsilon} &=& P_{\tilde{\delta}_h\tilde{\delta}_h}-P_{\delta_h\delta_h}=P_{\tilde{\delta}_h\tilde{\delta}_h}-b^2(k)P_{\delta\delta}\,.
\label{eq:Pshotnoise}
\eea
We compare the measurement of $P_{\epsilon\epsilon}$ with $1/\bar{n}_h$ of Poisson shot noise power spectrum in the right panel of Fig.~\ref{fig:shotnoise}. The measured stochastic terms show visible scale dependence even at linear scales. The scale dependent variation of $P_{\epsilon\epsilon}$ could reach $10\%$ in some cases, invalidating a constant parametrization for $P_{\epsilon\epsilon}$. Meanwhile, the $P_{\epsilon\epsilon}$ amplitudes of two halo mass bins show opposite trends. From $z=0.0$ to $z=1.5$, that of heavy bin increases from sub-Poissonian to super-Poissonian while that of light bin decreases from super-Poissonian to sub-Poissonian. Furthermore, both evolving trends are not purely monotonic.

Compared to a Poisson random point field, halo distribution consists of two extra properties, the halo exclusion and nonlinear clustering enhancement  outside the exclusion scale \cite{Baldauf13}. On one hand, halos have finite sizes. Within its radius, it is forbidden to randomly sample another halo from the dark matter density field. This halo exclusion breaks the Poisson assumption and causes a sub-Poissonian shot noise term. 
On the other hand, the nonlinear enhancement of clustering outside the exclusion scale will lead to a positive stochasticity correction of $P_{\epsilon\epsilon}$. The competition of these two factors results in the complicated $P_{\epsilon\epsilon}$ behavior as shown in figure~\ref{fig:shotnoise}. 

In our RSD model test, we will directly use the measured shot noise term for the fitting. The modelling of this stochastic term is beyond the scope of this paper. In realistic galaxy RSD analysis, the shot noise could be dramatically reduced by weighting halos with different masses \cite{Seljak09} or robustly modelled by proper fitting formula (e.g. \cite{Baldauf13,Zvonimir13,Vakili17,Ginzburg17}).

\subsubsection{Test of the halo density bias model}
\label{subsubsec:bias_test}

\begin{figure}[!t]
\centering
\includegraphics[width=0.49\columnwidth]{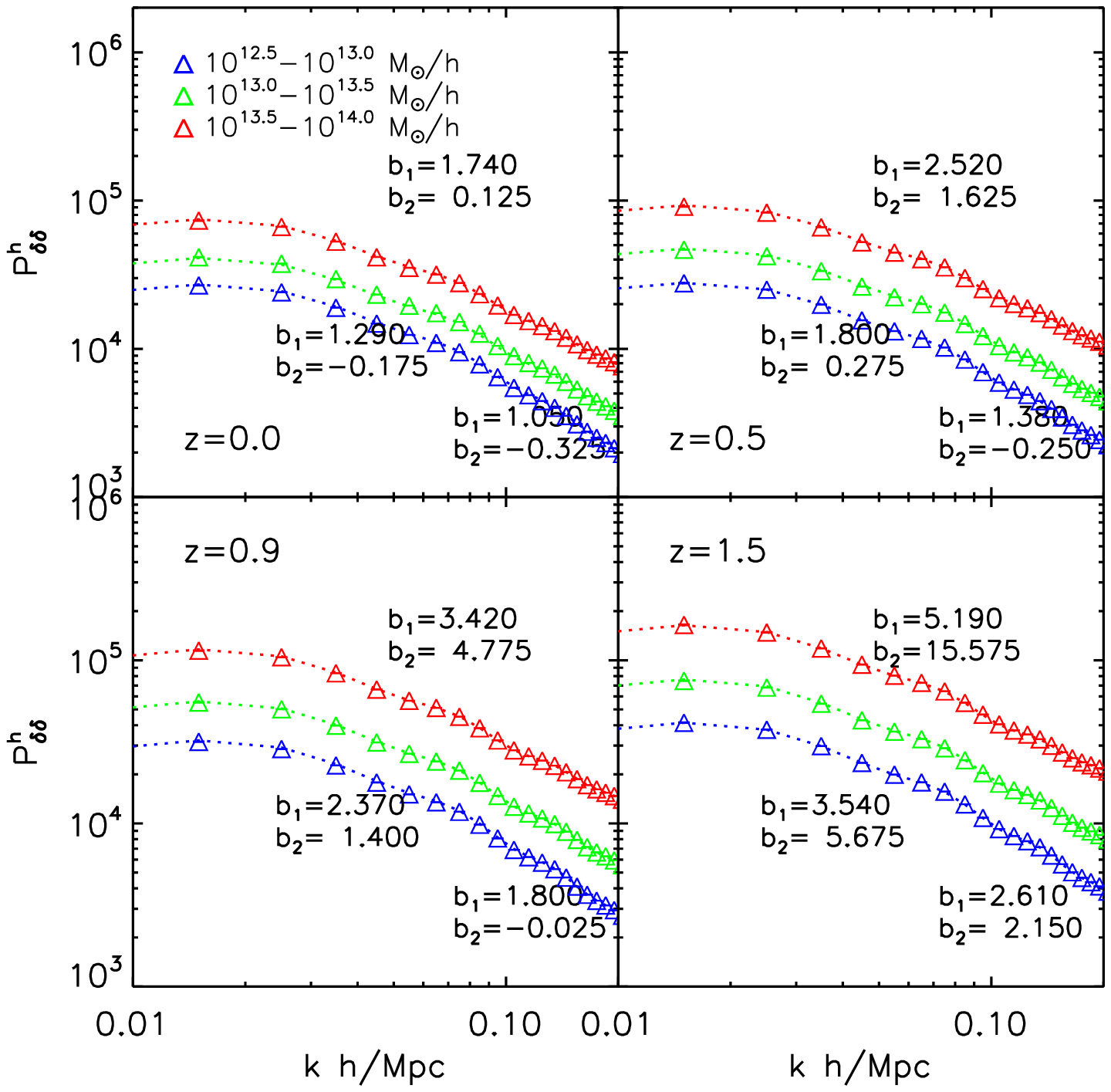}
\hfill
\includegraphics[width=0.49\columnwidth]{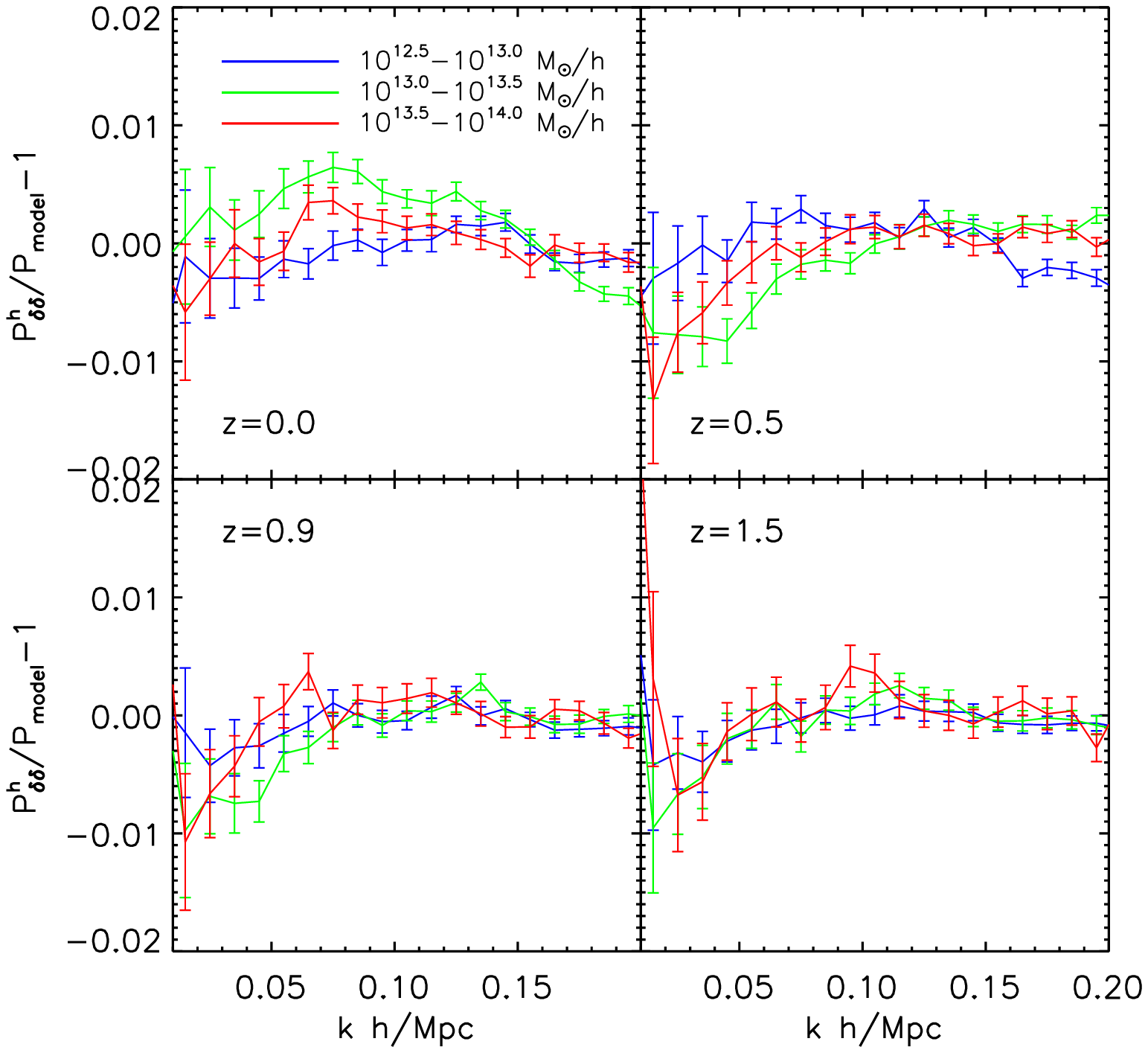}
\caption{Test of halo density bias model for predicting $P_{\delta_h\delta_h}$. {\it Left:} the triangles with error bars are measured $P_{\delta_h\delta_h}$. The dotted lines are fitted bias models from eq.~(\ref{eq:pdd_bias}) with corresponding fitted $b_1$ and $b_2$ also shown beside lines. {\it Right:} the fractional difference between the measurement and model.}
\label{fig:bias_phh}
\end{figure}

\begin{figure}[!t]
\centering
\includegraphics[width=0.49\columnwidth]{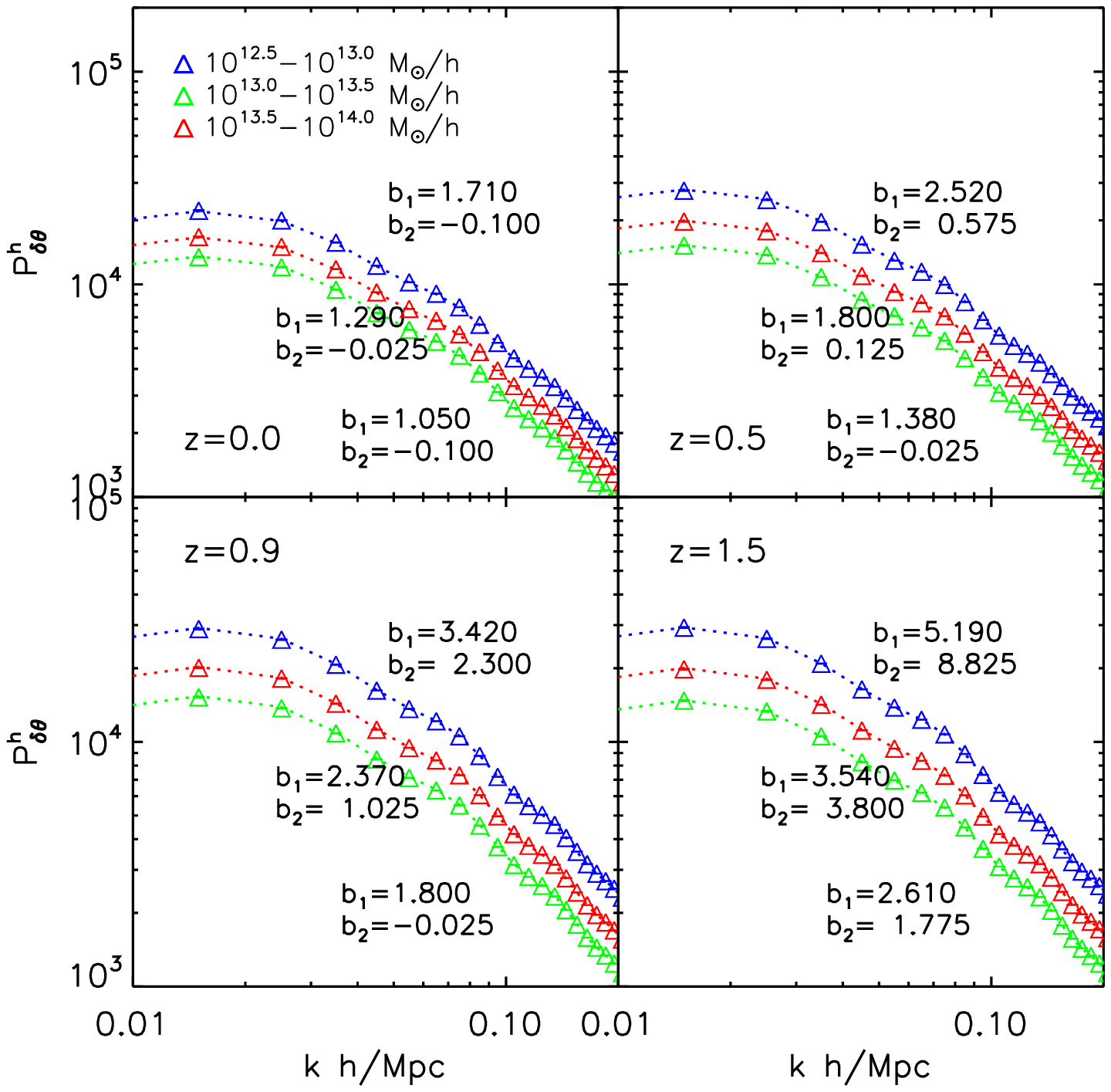}
\hfill
\includegraphics[width=0.49\columnwidth]{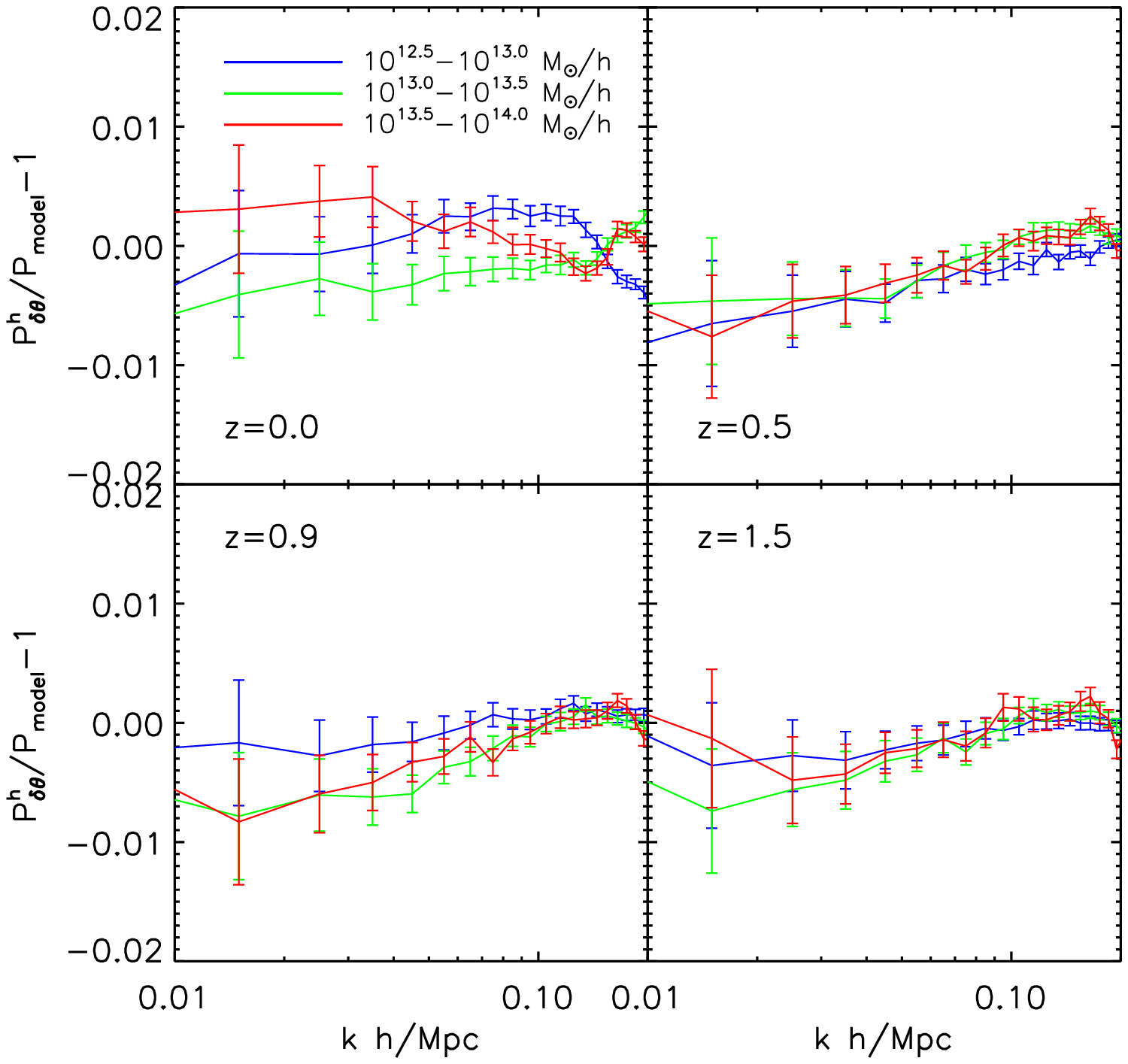}
\caption{Test of halo density bias model for predicting $P_{\delta_h\theta}$. {\it Left:} the triangles with error bars are measured $P_{\delta_h\theta}$. The dotted lines are fitted bias models from eq.~(\ref{eq:pdt_bias}) with corresponding fitted $b_1$ and $b_2$ also shown beside lines. {\it Right:} the fractional difference between the measurement and model.}
\label{fig:bias_pht}
\end{figure}

In this subsection, the accuracy of halo density bias model is tested in detail. The velocity bias keeps to be set $b_v=1$, and halo density fields are described by the model. Thus the cross--power spectrum $P_{\delta_h\theta_h}(k)$ is given by,
\beq
P_{\delta_h\theta_h}(k)=b_1P_{\delta\theta}(k)+b_2P_{b2,\theta}(k)+b_{s2}P_{bs2,\theta}(k)+b_{3\rm{nl}}\sigma_3^2(k)P^{\rm{L}}_{\rm m}(k)\,. 
\label{eq:pdt_bias2}
\eeq
where eq.~(\ref{eq:pdt_bias}) is applied. For the auto--power spectrum $P_{\delta_h\delta_h}$, we substitute eq.~(\ref{eq:phm_bias}) into eq.~(\ref{eq:pdd_bias}), which leads to,
\beq
P_{\delta_h\delta_h}(k)=\left(b_1P_{\delta\delta}(k)+b_2P_{b2,\delta}(k)+b_{s2}P_{bs2,\delta}(k)+b_{3\rm{nl}}\sigma_3^2(k)P^{\rm{L}}_{\rm m}(k) \right)^2/P_{\delta\delta}(k) \,. 
\label{eq:pdd_bias2}
\eeq
To be emphasized, here we are effectively modelling the cross--power spectrum $P_{\delta_h\delta}$, not directly the auto--power spectrum. We directly measure the $P_{\delta\theta}$ and $P_{\delta\delta}$ from simulations, and evaluate other power spectra in these two formulas by linear perturbation theory as described in Appendix \ref{appsec:highbias}. Then we fit two free parameters, $b_1$ and $b_2$, to test the accuracy of eqs.~(\ref{eq:pdd_bias2}) and~(\ref{eq:pdt_bias2}) in figures~\ref{fig:bias_phh} and~\ref{fig:bias_pht} respectively. In the left panels of figures, the triangles with error bars are simulation measurements, and the dotted lines are the fitted models. The fitted $b_1$ and $b_2$ are written besides the power spectra. In the right panels of the figures, the fractional differences between the measurements and models are plotted. We see that for all halo mass bins and redshifts, the adopted halo density bias model reaches $1\%$ accuracy and its uncertainty will not be a major systematic error in our RSD model test. Furthermore, the fitted $b_1$ from $P_{\delta_h\delta_h}$ and $P_{\delta_h\theta_h}$ are consistent with each other, showing the consistency of the halo density bias model.

\begin{figure}[!t]
\centering
\includegraphics[width=0.8\textwidth]{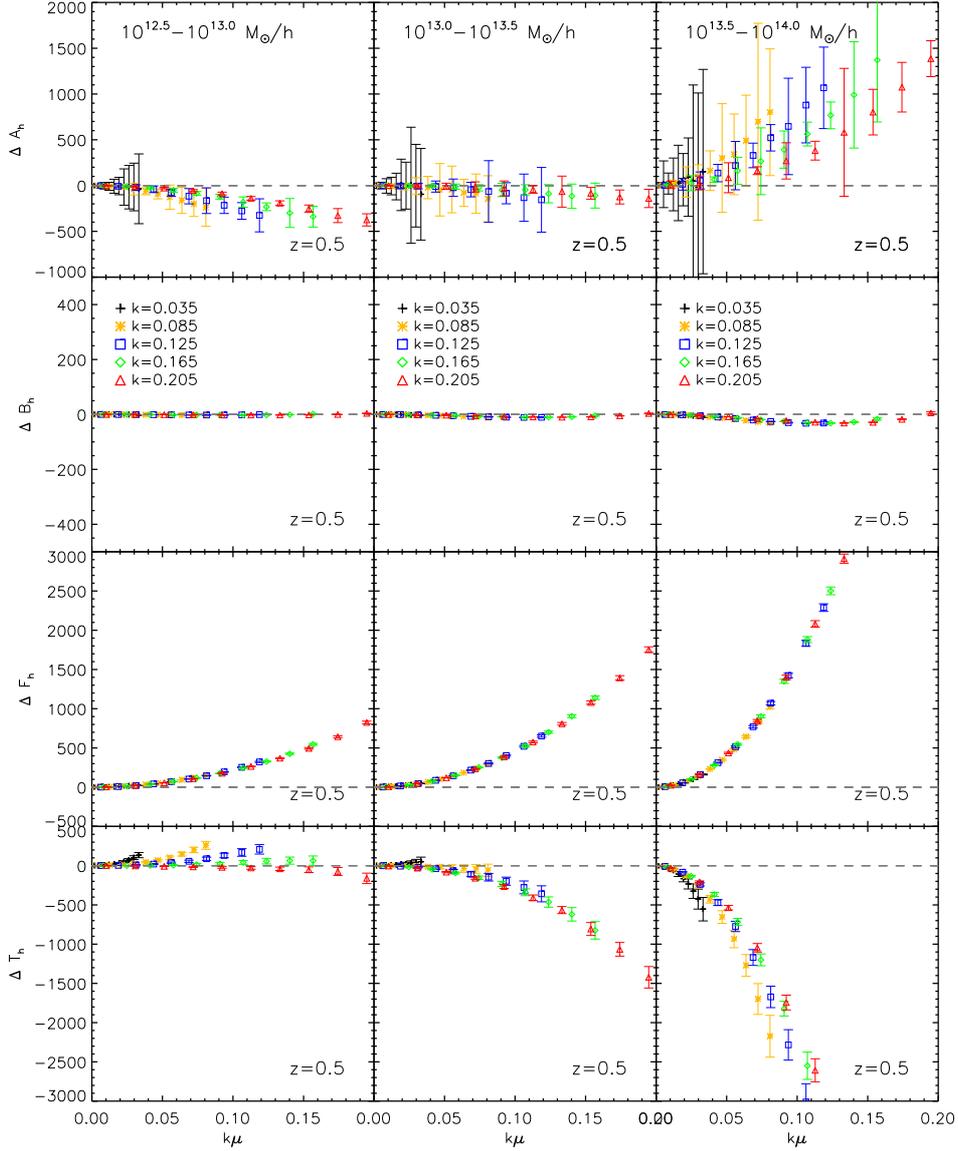}
\caption{From top to bottom, the differences between linear bias model predictions and direct measurements are respectively shown for $A_h$, $B_h$, $F_h$, $T_h$ terms for three halo mass bins. }
\label{fig:test_higher_diff}
\end{figure}

\begin{figure}[!t]
\centering
\includegraphics[width=0.8\textwidth]{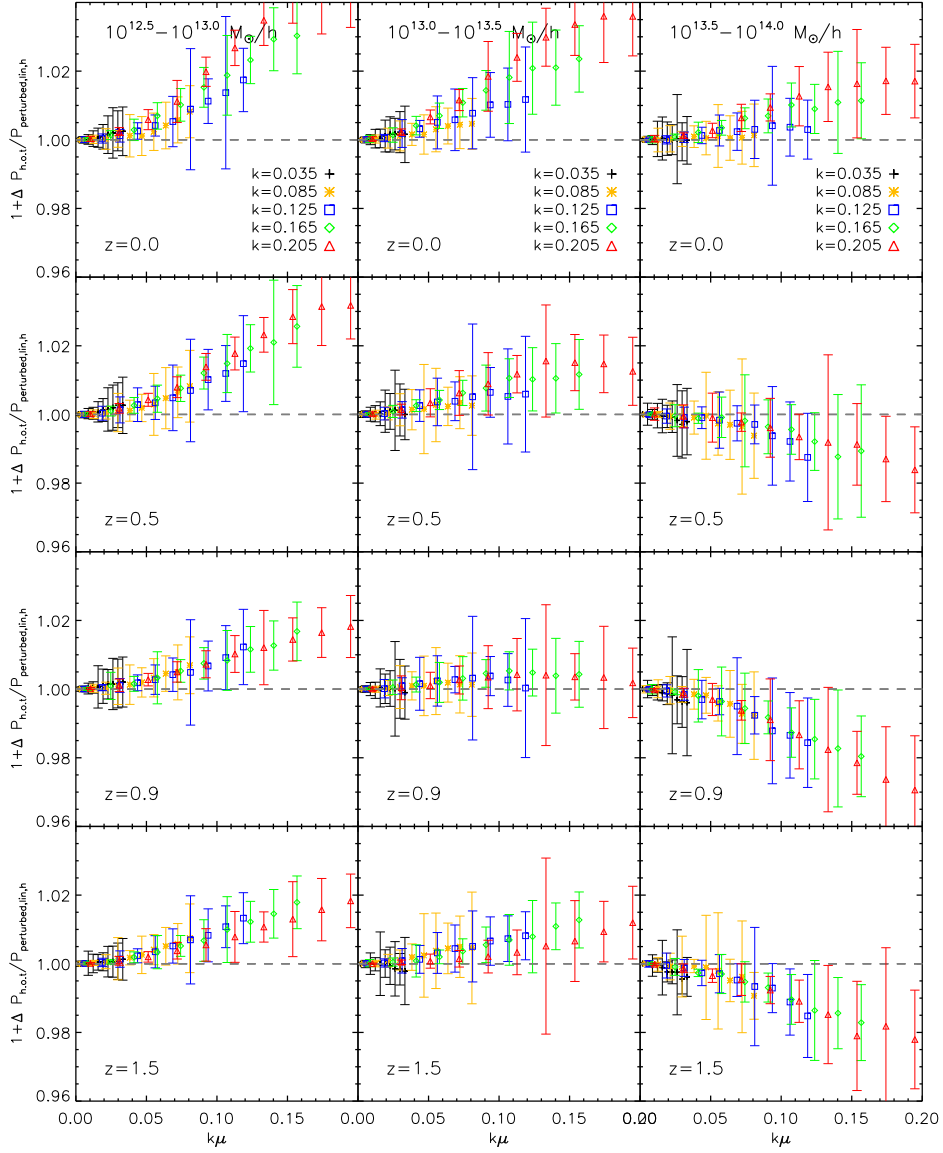}
\caption{The calculated $1+\Delta P_{\rm h.o.t}/P_{{\rm perturbed},h}$ of all halo mass bins and redshifts are shown.}
\label{fig:test_higher_ratio}
\end{figure}

\begin{figure}[!t]
\centering
\includegraphics[width=0.95\textwidth]{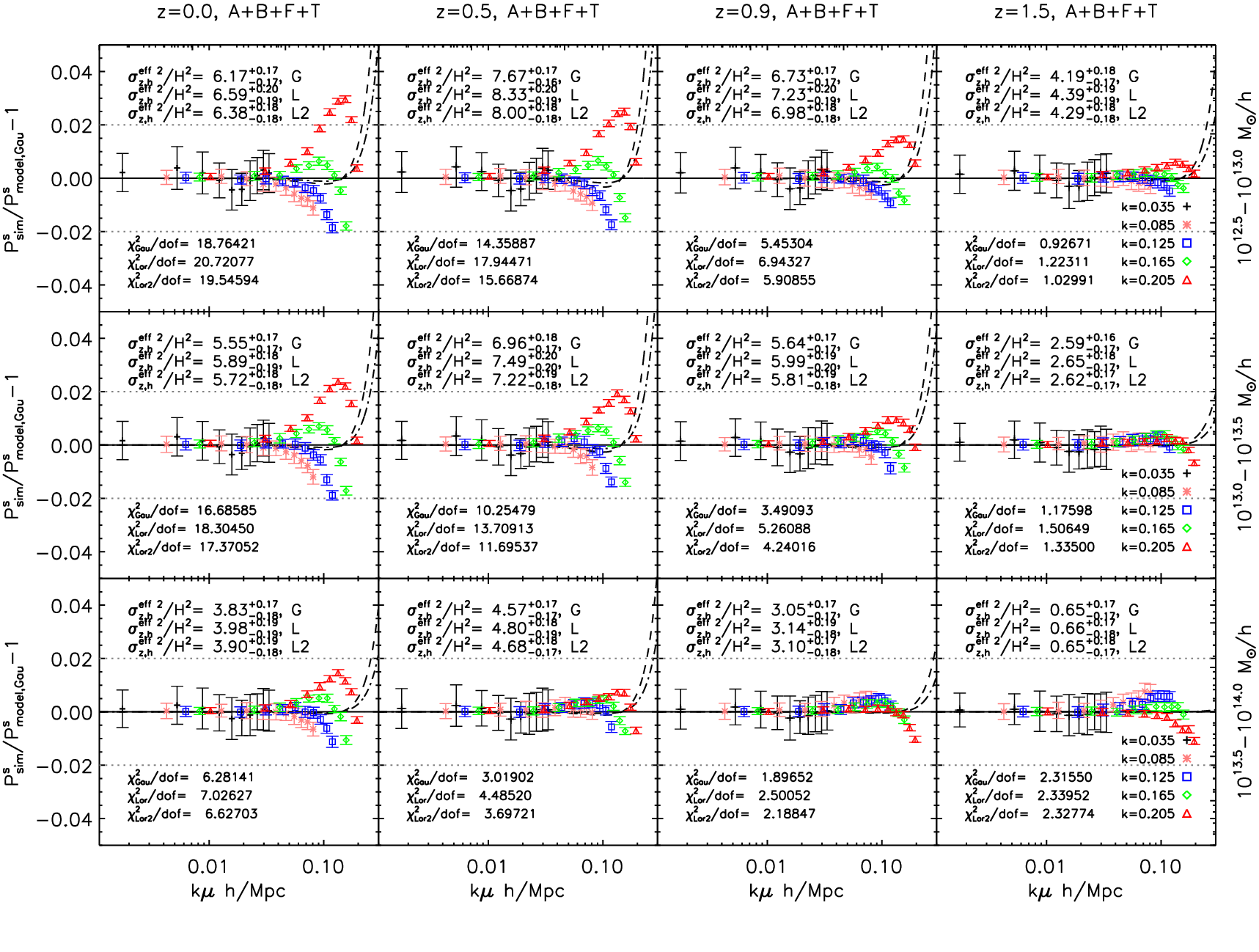}
\caption{Similar to figure~\ref{fig:fog_haloden_frac}, but during the fitting procedure, we calculate the higher order terms using linear bias model.}
\label{fig:fog_tnsb1}
\end{figure}

Next, the accuracy of higher order polynomial measurements under bias model are tested. It is still not clearly understood  what should be the precise bias models for higher order terms, particularly for $T_h$, the trispectrum-related one. Considering these higher order terms are not leading contributions in comparison to $P_{\delta_h\delta_h}$, $P_{\delta_h\theta_h}$  and $P_{\theta_h\theta_h}$, we simply approximate them with linear density bias here, as explained from eq.~(\ref{eq:higherorder_b1_A}) to eq.~(\ref{eq:higherorder_b1_T}) and in appendix~\ref{appsec:highbias} . In figure~\ref{fig:test_higher_diff}, from top to bottom, the differences between model predictions and measurements are respectively shown for $A_h$, $B_h$, $F_h$, $T_h$ terms at $z=0.5$. Except $B_h$ term, $A_h$, $F_h$, and $T_h$ terms could not be well predicted by the linear bias model. The differences are evident and the heavy halo mass bin presents larger deviations. Since heavy bin has more nonlinear density bias (e.g. larger fitted $b_2$ parameter), this deviation indicates the necessity of including higher order bias parameters in the modelling of higher order terms. Instead of implicitly including higher order bias parameters into calculation, we first check the model uncertainty of $A_h+B_h+F_h+T_h$ combination in figure~\ref{fig:test_higher_ratio}.

Here we define the model uncertainty of $A_h+B_h+F_h+T_h$ combination as $\Delta P_{\rm h.o.t}$ (``$\rm h.o.t$'' denotes ``higher order terms''),
\bea
\label{eq:Delta_hot}
\Delta P_{\rm h.o.t}&=&A_h(k,\mu)+B_h(k,\mu)+F_h(k,\mu)+T_h(k,\mu) \nonumber \\
&&-b_1^3A(k,\mu,f/b_1)-b_1^4B(k,\mu,f/b_1)-b_1^4F(k,\mu,f/b_1)-b_1^4T(k,\mu,f/b_1)
\eea
Inserting eq.~(\ref{eq:Delta_hot}) into eq.~(\ref{eq:Pkred_halo}), we have
\bea
P_h^{\rm (S)}(k,\mu)&=&D^{\rm FoG}(k\mu\sigma_{z,h})P_{{\rm perturbed},h}(k,\mu) \no \\
&=&D^{\rm FoG}(k\mu\sigma_{z,h})\left(P_{{\rm perturbed,lin},h}(k,\mu)+\Delta P_{\rm h.o.t}\right) \no \\
&=&D^{\rm FoG}(k\mu\sigma_{z,h})P_{{\rm perturbed,lin},h}(k,\mu)\left(1+\frac{\Delta P_{\rm h.o.t}}{P_{{\rm perturbed,lin},h}}\right)\no\\
&=&D^{\rm FoG}(k\mu\sigma^{\rm eff}_{z,h})P_{{\rm perturbed,lin},h}(k,\mu) \,,
\label{eq:Pkred_halo_higherorder}
\eea
where $P_{{\rm perturbed,lin},h}(k,\mu)$ denotes
\bea
P_{{\rm perturbed,lin},h}(k,\mu)&=&P_{\delta_h\delta_h}+2\mu^2P_{\delta_h\theta_h}+\mu^4P_{\theta_h\theta_h}  \\
&+&b_1^3A(k,\mu,f/b_1)+b_1^4B(k,\mu,f/b_1)+b_1^4F(k,\mu,f/b_1)+b_1^4T(k,\mu,f/b_1)\,. \no
\eea

The final step of eq.~(\ref{eq:Pkred_halo_higherorder}) shows that, the extra term $1+\Delta P_{\rm h.o.t}/P_{{\rm perturbed},h}$, induced by the inaccuracy of linear bias model in describing higher order terms, is absorbed into the FoG term and formulates an effective FoG term $D^{\rm FoG}(k\mu\sigma^{\rm eff}_{z,h})$. We plot the calculated $1+\Delta P_{\rm h.o.t}/P_{{\rm perturbed},h}$ of all halo mass bins and redshifts in figure~\ref{fig:test_higher_ratio}. Firstly these fractional ratios are shown to be simple functions of $k\mu$ within error bars, which coincides with our adopted FoG functional form. Secondly the fractional differences are relatively small, roughly within $3\%$, they could thus be well absorbed into the FoG term during fitting procedure and will not affect the model accuracy much. 

To further confirm the influence of linear bias model for higher order terms to the RSD model accuracy, we start to test the accuracy of eq.~(\ref{eq:Pkred_halo_higherorder}). The test is similar to that of eq.~(\ref{eq:Pkred_halo}), except that we replace the directly measured higher order terms with the linear bias model calculations. The results are shown in figure~\ref{fig:fog_tnsb1}. As expected, we see that the inaccuracy of linear bias model for higher order terms does not affect the RSD model accuracy much, and eq.~(\ref{eq:Pkred_halo_higherorder}) is shown to be accurate within $1\sim 2\%$ for all halo mass bins and redshifts.

Finally, the theoretical reason of the accidental cancellation of individual higher order term inaccuracy is interesting, together with the possibility of improving the individual higher order term model accuracy by including more higher order density bias parameters. Though beyond the scope of this paper, we will study these issues elsewhere.

\subsubsection{Test of the halo RSD model using full bias models}
\label{subsubsec:rsdtest_step3}

\begin{figure}[!t]
\centering
\includegraphics[width=0.95\textwidth]{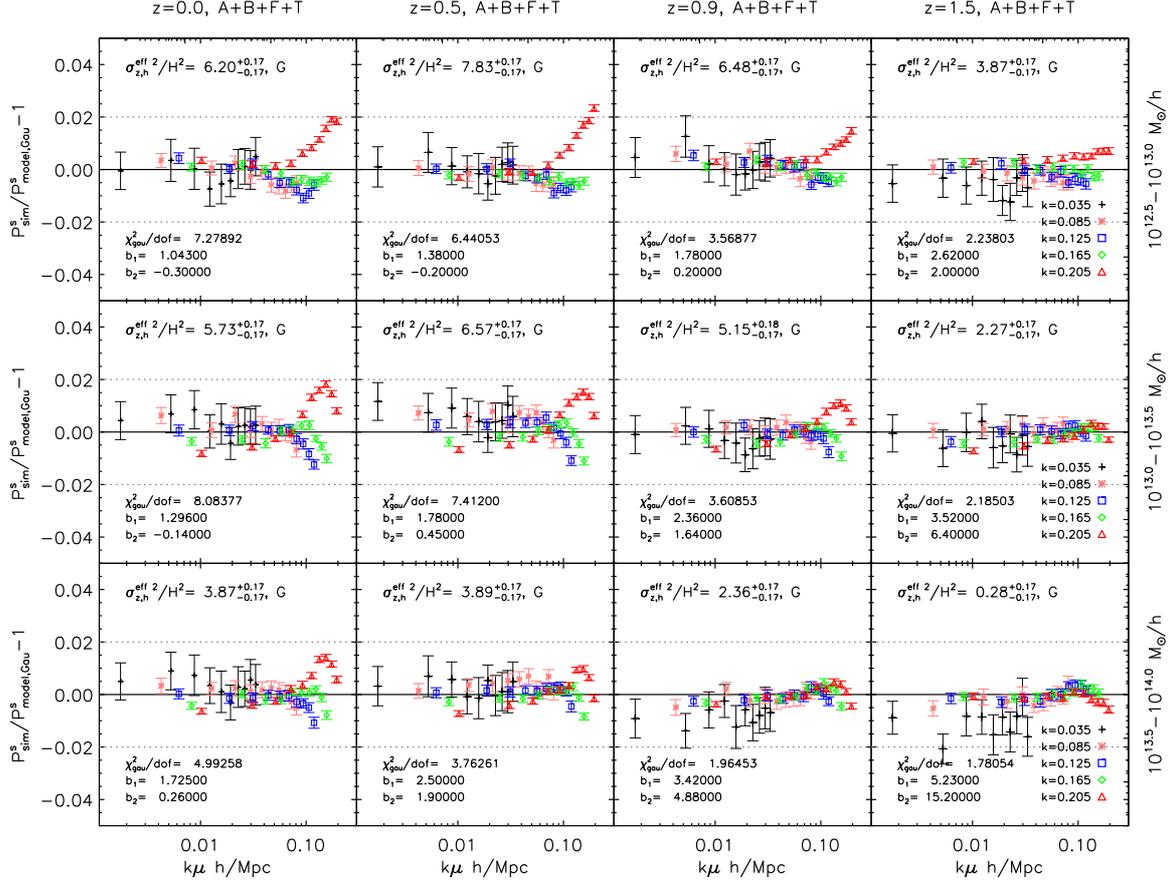}
\caption{Similar to figure~\ref{fig:fog_haloden_frac}, but we incorporate the full bias model into the halo mapping formula. We calculate real space $P_{\delta_h\delta_h}$, $P_{\delta_h\theta}$ and higher order terms using bias models in the fitting procedure.}
\label{fig:fog_full}
\end{figure}


The full halo RSD model is given by incorporating halo bias model which has been tested in the previous subsection as,
\bea
P_h^{\rm (S)}(k,\mu)&=&D^{\rm FoG}(k\mu\sigma^{\rm eff}_{z,h})[P_{\delta_h\delta_h}(k,b_1,b_2)+P_{\epsilon\epsilon}(k)+2\mu^2P_{\delta_h\theta}(k,b_1,b_2)+\mu^4P_{\theta\theta}(k) \no \\ 
 &+&b_1^3A(k,\mu,f/b_1)+b_1^4B(k,\mu,f/b_1)+b_1^4F(k,\mu,f/b_1)+b_1^4T(k,\mu,f/b_1)]\,.
\label{eq:Pkred_halo_fullbias}
\eea
Here we set $b_v=1$, and it is assumed that there is no uncertainty in the functional form of FoG given by the simple Gaussian function. When the underlying cosmology is known, there are three free parameters, $b_1$, $b_2$ and $\sigma^{\rm eff}_{z,h}$, to be varied to fit the theoretical model to the measurements.

The residual spectra for three halo mass bins are presented in the top, middle and bottom panels of figure~\ref{fig:fog_full} respectively. The reduced $\chi^2$ and the best fit $(b_1,b_2)$ are given at each panel. The tested results at different redshift of $z=(0.0,0.5,1.0,1.5)$ are shown from the first to fourth panels from the left. The dotted lines at each panel represent the tolerance of theoretical prediction up to 2\% level of accuracy. If the residual exceeds those bounds, then it indicates the failure of the theoretical halo RSD model. Shown in the figure, our model based upon eq.~(\ref{eq:Pkred_halo_fullbias}) accurately predicts the measurement within $1\sim2\%$ up to $k\lesssim0.2h$/Mpc, which is made possible by several contributions; (1) the accurate halo mapping formula of eq.~(\ref{eq:Pkred_halo}) verified in the section~\ref{subsec:rsd_model}, (2) the accurate halo density bias model of eq.~(\ref{eq:pdd_bias}) and eq.~(\ref{eq:pdt_bias}) explained in the section~\ref{subsubsec:bias_test}, (3) the absorption of several uncertainties from inaccurate bias model predictions into FoG term and $D^{\rm FoG}$ is well described by a Gaussian form, (4) the accurately measured dark matter templates (real space dark matter power spectra and higher order terms) from simulations.

We further discuss the above fourth point here. In RSD cosmology inference, the advantage of calculating dark matter templates from simulations rather than perturbation theory has been verified in \cite{Zheng16c,Song2018}. The improvement of numerical power will make our hybrid RSD model implementable in future data analysis.  In general, ``hybrid'' denotes the combination of simulation and theoretical calculation. The key spirit of our methodology is to search for the balance between these two and maximize the RSD model performance. Considering the model accuracy we have achieved, we expect that our hybrid RSD model will be a competitive data analysis tool for next generation dark energy projects. 

\section{Conclusion and discussions}
\label{sec:conclusion}

In this paper, we verify the accuracy of the halo RSD model~\cite{Taruya10,Zheng16a} which combines the advanced TNS formula \cite{Zheng16a} with halo bias model developed in~\cite{McDonald_bias}. The halo velocity bias is tested to be consistent with the unity at the targeted range of scale, which allows it to be hardwired $b_v=1$, with the verified fact that the influence of non-vanishing halo velocity bias at smaller scales on the halo power spectrum in redshift space could be absorbed into an effective Gaussian FoG term. The averaged measurement of halo anisotropic power spectra using 100 halo catalogs is exploited for the verification test. $A_h+B_h+F_h+T_h$, the complete higher order polynomial combination up to 2nd order of $k\mu$,  is used to compute the perturbative part of RSD model, and the FoG function is tested to be closed to Gaussian with only one free parameter of velocity dispersion. The real space dark matter templates in the perturbative part of model are computed using simulations rather than theoretical calculations, which makes the test immune from the uncertainty caused by perturbative description of non--linear physics. Halo clustering is constructed from dark matter clustering using both linear and non--linear biases. Three unknown parameters, FoG velocity dispersion $\sigma^{\rm eff}_{z,h}$, linear bias $b_1$ and non--linear bias $b_2$, are varied to fit the measured spectrum. We find that our model prediction is accurate within $1\sim 2\%$ at $k\lesssim0.2\hompc$ for all halo mass bins and redshifts.

While halo bias models work fine for two--point spectra like $P_{\delta_h\delta_h}$ and $P_{\delta_h\theta_h}$, the prediction for higher order polynomials does not work well. Since the non--linear bias modelling for higher order correlation functions is not known well, the linear bias model is solely used in this manuscript. This simple bias model does not predict correct individual higher order polynomial. Fortunately, the model uncertainties of higher order polynomials are canceled with each other, and the net effect becomes smaller. In addition, the pattern of this uncertainty in $k\mu$ space is consistent with FoG effect. Thus the effective FoG function is introduced to absorb this discrepancy. However, we would like to understand the bias model for higher order polynomials in a more rigorous way in our future work, by formulating the non--linear halo bias for these higher order polynomials.

We confirm the scale dependence of shot noise spectrum at linear scales in this work. While the direct measurement is used in this paper, a proper modelling of $P_{\epsilon\epsilon}$ will be necessary for the RSD analysis of next generation galaxy survey. Furthermore, $P_{\epsilon\epsilon}$ will be largely suppressed in the cross-power spectrum between different halo mass bins. It is interesting to verify our RSD model to this cross-power spectrum. We would like to address these issues in the future.

\section{Acknowledgments}
We thank the anonymous referee for carefully reading our manuscript and giving many insightful comments and suggestions. We thank Atsushi Taruya, Shun Saito, and Donghui Jeong for useful discussions. YZ thank Jeeson Song's help in generating the halo mock catalog. The work of running simulation was supported by the National Institute of Supercomputing and Network/Korea Institute of Science and Technology Information with supercomputing resources including technical support (KSC-2015-C1-017). Numerical calculations were performed by using a high performance computing cluster in the Korea Astronomy and Space Science Institute. To complete this work, discussions during the workshop, YITP-T-17-03, held at Yukawa Institute for Theoretical Physics (YITP) at Kyoto University were useful.

\appendix

\section{Higher order bias terms in power spectrum}
\label{appsec:highbias}

In this appendix, we present the detailed formulas to calculate the necessary parts of halo density bias model \cite{McDonald_bias}. We formulate $P_{\delta_h\delta_h}$ and $P_{\delta_h\theta_h}$ as
\bea
\label{eq:app_pdd_bias}
P_{\delta_h\delta_h}(k)&=&\left(b_1P_{\delta\delta}(k)+b_2P_{b2,\delta}(k)+b_{s2}P_{bs2,\delta}(k)+b_{3\rm{nl}}\sigma_3^2(k)P^{\rm{L}}_{\rm m}(k) \right)^2/P_{\delta\delta}(k) \,, \no \\
\label{eq:app_pdt_bias}
P_{\delta_h\theta_h}(k)&=&b_1P_{\delta\theta}(k)+b_2P_{b2,\theta}(k)+b_{s2}P_{bs2,\theta}(k)+b_{3\rm{nl}}\sigma_3^2(k)P^{\rm{L}}_{\rm m}(k)\,. \no
\eea
Here $P_{\delta\delta}$ and $P_{\delta\theta}$ are nonlinear dark matter power spectra, which are measured from simulation in this paper. $P^{\rm{L}}_{\rm m}$ is the linear dark matter power spectrum. We assume that the density bias is local in Lagrangian space. This implies  \cite{Baldauf2012,Chan2012,Saito2014}
\bea
b_{s2}=-\frac{4}{7}(b_1-1) \,,
&\quad&
b_{3\rm{nl}}=\frac{32}{315}(b_1-1)\,.\no
\eea
Three kernel functions are needed in the following formulations. They are generally expressed as
\bea
F_2(\bfk_1,\bfk_2)&=&\frac{5}{7}+\frac{1}{2}\frac{\bfk_1\cdot\bfk_2}{k_1k_2}\left(\frac{k_1}{k_2}+\frac{k_2}{k_1}\right)
+\frac{2}{7}\left[\frac{\bfk_1\cdot\bfk_2}{k_1k_2}\right]^2\,, \\
G_2(\bfk_1,\bfk_2)&=&\frac{3}{7}+\frac{1}{2}\frac{\bfk_1\cdot\bfk_2}{k_1k_2}\left(\frac{k_1}{k_2}+\frac{k_2}{k_1}\right)+\frac{4}{7}\left[\frac{\bfk_1\cdot\bfk_2}{k_1k_2}\right]^2\,, \\
S_2(\bfk_1,\bfk_2)&=&\left[\frac{\bfk_1\cdot\bfk_2}{k_1k_2}\right]^2-\frac{1}{3}\,.
\eea
The necessary power spectra in eq.~(\ref{eq:app_pdd_bias}) are calculated by
\bea
\label{eq:pb2d}
P_{b2,\delta}&=&\int \frac{d^3q}{(2\pi)^3}\,P^{\rm L}_{\rm m}(q)P^{\rm L}_{\rm m}(|\bfk-\bfq|)F_2(\bfq,\bfk-\bfq) \nonumber \\
&=&\frac{1}{(2\pi)^2}\int dqd\mu\, q^2P^{\rm L}_{\rm m}(q)P^{\rm L}_{\rm m}(\sqrt{k^2-2qk\mu+q^2})F^{(2)}_{\rm S}(\bfq,\bfk-\bfq)\,, \\
\label{eq:pbs2d}
P_{bs2,\delta}&=&\int \frac{d^3q}{(2\pi)^3}\,P^{\rm L}_{\rm m}(q)P^{\rm L}_{\rm m}(|\bfk-\bfq|)F_2(\bfq,\bfk-\bfq)S_2(\bfq,\bfk-\bfq) \,, \nonumber \\
&=&\frac{1}{(2\pi)^2}\int dqd\mu\, q^2P^{\rm L}_{\rm m}(q)P^{\rm L}_{\rm m}(\sqrt{k^2-2qk\mu+q^2})F_2(\bfq,\bfk-\bfq)S_2(\bfq,\bfk-\bfq) \,,\\
\label{eq:sigma32}
\sigma_3^2(k)&=&\int \frac{d^3q}{(2\pi)^3}\,P^{\rm L}_{\rm m}(q)\left[\frac{5}{6}+\frac{15}{8}S_2(\bfq,\bfk-\bfq)S_2(-\bfq,\bfk)-\frac{5}{4}S_2(\bfq,\bfk-\bfq)\right] \nonumber \\
&=&\frac{1}{(2\pi)^2}\int dqd\mu\, q^2P^{\rm L}_{\rm m}(q)\left[\frac{5}{6}+\frac{15}{8}S_2(\bfq,\bfk-\bfq)S_2(-\bfq,\bfk)-\frac{5}{4}S_2(\bfq,\bfk-\bfq)\right]\,,
\eea
in which we have kernels
\bea
F_2(\bfq,\bfk-\bfq)&=&\frac{5}{7}+\frac{1}{2}\frac{qk\mu-q^2}{q\sqrt{k^2-2qk\mu+q^2}}\left(\frac{q}{\sqrt{k^2-2qk\mu+q^2}}+\frac{\sqrt{k^2-2qk\mu+q^2}}{q}\right) \no\\
&&+\frac{2}{7}\left[\frac{qk\mu-q^2}{q\sqrt{k^2-2qk\mu+q^2}}\right]^2\,, \nonumber\\
S_2(\bfq,\bfk-\bfq)&=&\left[\frac{qk\mu-q^2}{q\sqrt{k^2-2qk\mu+q^2}}\right]^2-\frac{1}{3}\,.\nonumber\\
S_2(-\bfq,\bfk)&=&\left[\frac{-qk\mu}{qk}\right]^2-\frac{1}{3}\,. \no
\eea
Similarly, the necessary power spectra in eq.~(\ref{eq:app_pdt_bias}) are calculated by
\bea
\label{eq:pb2t}
P_{b2,\theta}&=&\int \frac{d^3q}{(2\pi)^3}\,P^{\rm L}_{\rm m}(q)P^{\rm L}_{\rm m}(|\bfk-\bfq|)G_2(\bfq,\bfk-\bfq) \nonumber \\
&=&\frac{1}{(2\pi)^2}\int dqd\mu\, q^2P^{\rm L}_{\rm m}(q)P^{\rm L}_{\rm m}(\sqrt{k^2-2qk\mu+q^2})G_2(\bfq,\bfk-\bfq)\,, \\
\label{eq:pbs2t}
P_{bs2,\theta}&=&\int \frac{d^3q}{(2\pi)^3}\,P^{\rm L}_{\rm m}(q)P^{\rm L}_{\rm m}(|\bfk-\bfq|)G_2(\bfq,\bfk-\bfq)S_2(\bfq,\bfk-\bfq) \nonumber \\
&=&\frac{1}{(2\pi)^2}\int dqd\mu\, q^2P^{\rm L}_{\rm m}(q)P^{\rm L}_{\rm m}(\sqrt{k^2-2qk\mu+q^2})G_2(\bfq,\bfk-\bfq)S_2(\bfq,\bfk-\bfq) \,,
\eea
in which
\bea
G_2(\bfq,\bfk-\bfq)&=&\frac{3}{7}+\frac{1}{2}\frac{qk\mu-q^2}{q\sqrt{k^2-2qk\mu+q^2}}\left(\frac{q}{\sqrt{k^2-2qk\mu+q^2}}+\frac{\sqrt{k^2-2qk\mu+q^2}}{q}\right) \no\\
&&+\frac{4}{7}\left[\frac{qk\mu-q^2}{q\sqrt{k^2-2qk\mu+q^2}}\right]^2\,. \nonumber
\eea

\section{Higher order polynomial calculations}
\label{appsec:ABFT}

We present the details of higher order term calculation using linear bias model in this appendix. We assume $b_v=1$. In linear density bias model, we have $\delta_h(\bfk)=b_1\delta(\bfk)$, $\delta_h(\bfx)=b_1\delta(\bfx)$, thus 
\bea
\label{eq:A_h}
A_h(k,\mu)&=& j_1\,\int d^3\bfx \,\,e^{i\bfk\cdot\bfx}\,\,\langle A_1A_2A_3\rangle_c\nonumber\\
&=&j_1\,\int d^3\bfx \,\,e^{i\bfk\cdot\bfx}\,\,\langle (u_{z,h}-u'_{z,h})(\delta_h+\nabla_zu_{z,h})(\delta'_h+\nabla_zu'_{z,h})\rangle_c\, \\
&\simeq &j_1\,\left[b_1^2\int d^3\bfx \,\,e^{i\bfk\cdot\bfx}\,\,\langle u_z\delta\delta'\rangle_c+b_1\int d^3\bfx \,\,e^{i\bfk\cdot\bfx}\,\,\langle u_z\delta\nabla_zu'_z\rangle_c \right.\nonumber\\
&&+b_1\int d^3\bfx \,\,e^{i\bfk\cdot\bfx}\,\,\langle u_z\nabla_z u_z\delta'\rangle_c+\int d^3\bfx \,\,e^{i\bfk\cdot\bfx}\,\,\langle u_z\nabla_z u_z\nabla_z u'_z\rangle_c \nonumber\\
&&-b_1^2\int d^3\bfx \,\,e^{i\bfk\cdot\bfx}\,\,\langle \delta u_z'\delta'\rangle_c- b_1\int d^3\bfx \,\,e^{i\bfk\cdot\bfx}\,\,\langle \delta u_z'\nabla_zu_z'\rangle_c \nonumber\\
&&\left.-b_1\int d^3\bfx \,\,e^{i\bfk\cdot\bfx}\,\,\langle \nabla_zu_zu_z'\delta'\rangle_c-\int d^3\bfx \,\,e^{i\bfk\cdot\bfx}\,\,\langle \nabla_zu_z u'_z\nabla_z u'_z\rangle_c \right] \,.
\eea
In a similar way, $B(k,\mu)$ could be expressed as
\bea
\label{eq:B_h}
B_h(k,\mu)&=& j_1^2\,\int d^3\bfx \,\,e^{i\bfk\cdot\bfx}\,\,\langle A_1A_2\rangle_c\,\langle A_1A_3\rangle_c\nonumber\\
&=&j_1^2\,\int d^3\bfx \,\,e^{i\bfk\cdot\bfx}\,\,\langle (u_{z,h}-u'_{z,h})(\delta_h+\nabla_zu_{z,h})\rangle_c\langle (u_{z,h}-u'_{z,h})(\delta'_h+\nabla_zu'_{z,h})\rangle_c \no \\
&\simeq& -j_1^2\left[b_1^2\int d^3\bfx \,\,e^{i\bfk\cdot\bfx}\,\,\langle \delta u'_z\rangle_c\langle u_z\delta'\rangle_c+b_1\int d^3\bfx \,\,e^{i\bfk\cdot\bfx}\,\,\langle \nabla_zu_{z} u'_z\rangle_c\langle u_z\delta'\rangle_c \right.\nonumber\\
&&+\left.b_1\int d^3\bfx \,\,e^{i\bfk\cdot\bfx}\,\,\langle \delta u'_z\rangle_c\langle u_z\nabla_zu'_z\rangle_c+\int d^3\bfx \,\,e^{i\bfk\cdot\bfx}\,\,\langle \nabla_zu_{z} u'_z\rangle_c\langle u_z\nabla_zu'_z\rangle_c \right]\,.
\eea
$F(k,\mu)$ could be expressed as
\bea
\label{eq:F_h}
F_h(k,\mu)&=& -j_1^2\,\int d^3\bfx \,\,e^{i\bfk\cdot\bfx}\,\,\langle u_{z,h} u_{z,h}'\rangle_c\langle A_2A_3\rangle_c \nonumber\\
&=&-j_1^2\,\int d^3\bfx \,\,e^{i\bfk\cdot\bfx}\,\,\langle u_{z,h} u_{z,h}'\rangle_c \langle (\delta_h+\nabla_zu_{z,h})(\delta'_h+\nabla_zu_{z,h}')\rangle_c \no \\
&\simeq& -j_1^2\left[b_1^2\int d^3\bfx \,\,e^{i\bfk\cdot\bfx}\,\,\langle u_z u'_z\rangle_c\langle \delta\delta'\rangle_c+b_1\int d^3\bfx \,\,e^{i\bfk\cdot\bfx}\,\,\langle u_z u'_z\rangle_c\langle \delta\nabla_zu_z'\rangle_c \right.\nonumber\\
&&+\left.b_1\int d^3\bfx \,\,e^{i\bfk\cdot\bfx}\,\,\langle u_z u'_z\rangle_c\langle \nabla_zu_z\delta'\rangle_c+\int d^3\bfx \,\,e^{i\bfk\cdot\bfx}\,\,\langle u_z u'_z\rangle_c\langle \nabla_zu_z\nabla_zu'_z\rangle_c \right]\,.
\eea
$T(k,\mu)$ could be expressed as  
\bea
\label{eq:T_h}
T(k,\mu)&=& \frac{1}{2} j_1^2\,\int d^3\bfx \,\,e^{i\bfk\cdot\bfx}\,\,\langle A_1^2A_2A_3\rangle_c \nonumber \\
&=&\frac{1}{2} j_1^2\,\int d^3\bfx \,\,e^{i\bfk\cdot\bfx}\,\,\left\{\langle (u_{z,h}-u_{z,h}')^2 (\delta_h+\nabla_zu_{z,h})(\delta_h'+\nabla_zu_{z,h}')\rangle_c \right.\no \\
&&\left.-\langle A_1^2\rangle\langle A_2A_3\rangle-2\langle A_1A_2\rangle\langle A_1A_3\rangle\right\} \no \\
&\simeq&\frac{1}{2}j_1^2\left[b_1^2\int d^3\bfx \,\,e^{i\bfk\cdot\bfx}\langle u_z^2\delta\delta'\rangle + b_1 \int d^3\bfx \,\,e^{i\bfk\cdot\bfx}\langle u_z^2\delta\nabla_zu_z'\rangle\right. \nonumber\\
&&+b_1\int d^3\bfx \,\,e^{i\bfk\cdot\bfx}\langle u_z^2\nabla_zu_z\delta'\rangle + \int d^3\bfx \,\,e^{i\bfk\cdot\bfx}\langle u_z^2\nabla_zu_z\nabla_zu_z'\rangle \nonumber \\
&&-2b_1^2\int d^3\bfx \,\,e^{i\bfk\cdot\bfx}\langle u_z\delta u_z'\delta'\rangle - 2b_1 \int d^3\bfx \,\,e^{i\bfk\cdot\bfx}\langle u_z\delta u_z'\nabla_zu_z'\rangle\nonumber\\
&&-2b_1\int d^3\bfx \,\,e^{i\bfk\cdot\bfx}\langle u_z\nabla_zu_z u_z'\delta'\rangle - 2\int d^3\bfx \,\,e^{i\bfk\cdot\bfx}\langle u_z\nabla_zu_zu_z'\nabla_zu_z'\rangle \nonumber \\
&&+b_1^2\int d^3\bfx \,\,e^{i\bfk\cdot\bfx}\langle \delta u_z'^2\delta'\rangle + b_1 \int d^3\bfx \,\,e^{i\bfk\cdot\bfx}\langle \delta u_z'^2\nabla_zu_z'\rangle \nonumber\\
&&\left.+b_1\int d^3\bfx \,\,e^{i\bfk\cdot\bfx}\langle \nabla_zu_zu_z'^2\delta'\rangle + \int d^3\bfx \,\,e^{i\bfk\cdot\bfx}\langle \nabla_zu_zu_z'^2\nabla_zu_z'\rangle \right] \no\\
&&-B_h(k,\mu,b_1)-F_h(k,\mu,b_1)\no\\
&&-j_1^2\sigma_{z,h}^2\left[P_{\delta_h\delta_h}(k,b_1,b_2)+2\mu^2P_{\delta_h\theta_h}(k,b_1,b_2,b_v)+\mu^4P_{\theta_h\theta_h}(k,b_1,b_2,b_v) \right] \,.
\eea
Here $\sigma_{z,h}^2$ is directly measured from the sampled dark matter velocity field, rather than treated as a free parameter.

\bibliographystyle{JHEP}
\bibliography{mybib}
\end{document}